\documentclass[12pt]{article}
\usepackage[left=1.1in, right=1.1in, top=27mm, bottom=35mm]{geometry}
\usepackage{url}
\usepackage{amsmath,amssymb,amsfonts}
\newcommand{\tx}{\text}
\usepackage{hyperref}
\usepackage{float}
\usepackage{hyperref}
\usepackage{graphicx}
\usepackage{listings}
\usepackage{xcolor}
\usepackage{mathrsfs}
\usepackage{titling,lipsum}
\usepackage{ragged2e} 
\usepackage{tikz}
\usetikzlibrary{matrix}
\usetikzlibrary{tikzmark}
\usepackage{cite}

\definecolor{codegreen}{rgb}{0,0.6,0}
\definecolor{codegray}{rgb}{0.5,0.5,0.5}
\definecolor{codepurple}{rgb}{0.58,0,0.82}
\definecolor{backcolour}{rgb}{1,1,1}

\begin{document}

\begin{titlingpage}

\begin{flushright}
QMUL-PH-23-32\\
\end{flushright}

\vspace{1cm}

\centering
{\scshape\LARGE  \textbf{Constructing and Machine Learning \\ Calabi-Yau Five-folds}\par}

\vspace{0.75cm}
{}\par

\vspace{0.75cm}

{R. Alawadhi$^a$, D. Angella$^b$, A. Leonardo$^b$ and T. Schettini Gherardini$^a$ \par
\vspace{1cm} 
$^a$Centre for Theoretical Physics, School of Physical and Chemical Sciences \\
Queen Mary University of London, 327 Mile End
Road, London E1 4NS, UK \\
$^b$ Dipartimento di Matematica e Informatica "Ulisse Dini", Universita' degli Studi di Firenze, Viale Morgagni 67/A, 50134 Firenze, Italy \\ \vspace{1cm}
E-mail:  \href{mailto:r.alawadhi@qmul.ac.uk}{r.alawadhi@qmul.ac.uk}, \href{mailto:daniele.angella@unifi.it}{daniele.angella@unifi.it}, \href{mailto:andrea.leonardo@edu.unifi.it}{andrea.leonardo@edu.unifi.it}, \href{mailto:t.schettinigherardini@qmul.ac.uk}{t.schettinigherardini@qmul.ac.uk} }

\vspace{2cm}

\justifying
{ABSTRACT: Motivated by their role in M-theory, F-theory and S-theory compactifications, we construct all possible complete intersection Calabi-Yau five-folds in a product of four or less complex projective spaces, with up to four constraints. We obtain $27068$ spaces, which are not related by permutations of rows and columns of the configuration matrix, and determine the Euler number for all of them. Excluding the $3909$ product manifolds among those, we calculate the cohomological data for $12433$ cases, i.e. $53.7 \%$ of the non-product spaces, obtaining $2375$ different Hodge diamonds. The dataset containing all the above information is available \href{https://www.dropbox.com/scl/fo/z7ii5idt6qxu36e0b8azq/h?rlkey=0qfhx3tykytduobpld510gsfy&dl=0}{\textcolor{blue}{here}}. The distributions of the invariants are presented, and a comparison with the lower-dimensional analogues is discussed. Supervised machine learning is performed on the cohomological data, via classifier and regressor (both fully connected and convolutional) neural networks. We find that $h^{1,1}$ can be learnt very efficiently, with very high $R^2$ score and an accuracy of $96\%$, i.e. $96 \%$ of the predictions exactly match the correct values. For $h^{1,4},h^{2,3}, \eta$, we also find very high $R^2$ scores, but the accuracy is lower, due to the large ranges of possible values. 
}

\end{titlingpage}

\tableofcontents

\newpage

\section{Introduction}
Calabi-Yau manifolds are of paramount importance in string theory, since they (together with their orbifolds) are the most promising candidates for internal spaces in the compactification mechanism \cite{Candelas:1985en}.
The simplest example of a Calabi-Yau manifold arises when considering the zero locus of a homogeneous polynomial in the complex projective space. The natural generalisation of this setup consists of having a number of homogeneous polynomials living in a product of projective spaces; under some specific conditions, the zero locus of such polynomials defines a Calabi-Yau manifold. Spaces obtained using the above prescription are usually referred as complete intersection Calabi-Yau's, or CICY's for short. This algebro-geometric construction, first introduced in the seminal work of Candelas \cite{CANDELAS1988493}, was a fundamental result, since it provided a systematic way of producing examples of Calabi-Yau manifolds of complex dimension three. CICY three-folds were classified (up to some equivalences, see \cite{Anderson_2008}) in \cite{Green:1987cr}, counting $265$ distinct Hodge diamonds among roughly $8000$ manifolds. A number of other techniques were also devised to produce examples of Calabi-Yau manifolds \cite{CANDELAS1990383, kreuzer2000complete, SKARKE_1996}, motivating the community to collect large - sometimes extremely large - sets of data pertaining to such spaces, arranged in massive databases of algebro-geometric information (see \cite{KSdata}, for instance). These lists focus mainly on the Hodge numbers, since they play the roles of physical parameters in the lower-dimensional theory obtained via compactification \cite{Candelas:1985en, Bull:2018uow}. The importance of such holomorphic invariants, also in the context of pure mathematics, cannot be underestimated: they count the number of classes of K\"ahler metrics and the possible complex deformations, just to mention the two most quoted results. In the age of big data and increased computing power, the databases mentioned above also provide a fertile land for machine learning to be effective. In fact, employing neural networks to the study of CICY three-folds' Hodge numbers was one of the earliest applications of machine learning to theoretical physics, as was presented in \cite{HE2017564}.  A number of papers also investigating neural network performances on Calabi-Yau three-folds datasets followed in the subsequent years \cite{Bull:2018uow, Bull:2019cij, Erbin_2021, Berman_2022}.

While three-folds are suitable candidates for compactifications of theories with ten dimensions, their higher-dimensional analogues need to be considered when dealing with theories having more than ten dimensions. M-theory \cite{Horava:1995qa, Lukas:1998tt, EmilianDudas_2000}, F-theory \cite{Vafa:1996xn, Morrison:1996na, Sen:1996vd, Weigand:2018rez}, and S-theory \cite{Bars:1996ab} are the most famous examples of such theories.
Motivated by the study of F-theory vacua (see \cite{Gukov:1999ya,Grimm:2011sk}), Gray, Haupt and Lukas classified all CICY four-folds, obtaining more than $900000$ inequivalent configurations, with over $4000$ different sets of Hodge numbers, in \cite{Gray_2013, Gray_2014}. The dataset was investigated via different machine learning techniques in \cite{He:2020lbz, Erbin_2022} with very promising outcomes.  More recently, five-folds also made an appearance in the physics literature, although studies specifically dedicated to classifying Calabi-Yau five-folds are scarce (we could only find a partial classification in \cite{KSdata}, with incomplete Hodge information). M-theory, when compactified on a Calabi-Yau five-fold, results in a $\mathcal{N}=2$ supersymmetric quantum mechanics \cite{Haupt:2008nu, haupt2009mtheory}. Moreover, Calabi-Yau five-folds play a role in F-theory, where upon compactification, they provide a way to systematically construct $N = (0,2)$ CFTs \cite{Schafer-Nameki:2016cfr, Tian:2020yex}, which might allow to classify them. Therefore, a good dataset of such Calabi-Yau manifolds is crucial in the search of relevant CFTs. Additionally, in \cite{Curio:1998bv}, a 3D string vacua has been constructed by compactifying S-theory on a $T^2\times T^3$ fibered Calabi-Yau five-fold. 

Having outlined the context for the present work from the string theory perspective, let us review some recent developments in the application of state-of-the-art data science techniques in this field. As mentioned above, machine learning as a tool for exploring the landscape of string compactifications on Calabi-Yau manifolds has appeared multiple times in the recent literature (see \cite{He:2018jtw, Carifio:2017bov, Ruehle:2020jrk, Demirtas:2020dbm, RUEHLE20201,Erbin_2021_m, Constantin:2021for, Abel:2021ddu,Aggarwal:2023swe, Berglund:2023ztk, Bao:2021ofk, Bao:2020sqg}).\footnote{Some further applications of machine learning to theoretical physics include approximating metrics \cite{Ashmore:2019wzb, Larfors:2022nep, Cui:2019uhy, Anderson:2020hux}, analysing CFTs and the AdS/CFT correspondence \cite{PhysRevLett.128.041601, Kantor:2021jpz, Hashimoto:2018ftp, Hashimoto:2019bih}, and investigating lattice QCD \cite{PhysRevD.103.014509, Yoon:2018krb, Boyda:2022nmh}, just to name a few.} These investigations uncovered new features and insights, such as an unobserved clustering behaviour of the Hodge numbers \cite{Berman_2022}, a new approximation method \cite{Ed}, and  hints for an unknown formula for $h^{1,3}$ \cite{He:2020lbz}. Restricting to CICY's only, astonishingly good results with $100\%$ accuracy have been obtained for four-folds in \cite{Erbin_2022}, and investigations in \cite{Erbin_2023} show new promising results on extrapolating predictions from low to high Hodge numbers for both three-folds and four-folds. The latter technique is particularly relevant for the case of five-folds, whose exhaustive list is estimated to be astronomical in size; such an approach might allow to extract global properties of the dataset quickly without the need to generate it all (which could be unfeasible with current computational resources).

Our work aims to be the first useful effort in classifying CICY five-folds and applying machine learning techniques to the resulting dataset. From the sizes of the CICY three-folds and four-folds datasets, we expect the five-folds one to be of the order of $10^8$. Given the computational power at our disposal, we limit ourselves to a subset of it for this first investigation.
 
The paper is structured as follows. In section \ref{sec:2} we review the mathematics behind CICY five-folds. We start by recalling general facts about Calabi-Yau manifolds and outlining the construction of such manifolds as complete intersections (sections \ref{sec:The_construction}
and \ref{sec:Basic_properties}). Then, section \ref{sec:Spectral_sequences} presents the techniques that were employed for the calculation of the Hodge numbers. We build upon the analogous works in the context of three-folds and four-folds (summarised in \cite{Green:1987cr, Gray_2014}), which are adapted to the case under investigation in sub-section \ref{sec:Adjunction}; but we also introduce a further and new step in the computation, described in sub-section \ref{sec:Sym_adjunction}, which is needed to determine the full Hodge diamond. Section \ref{sec:Building_dataset} is devoted to outlining how the dataset was constructed. We comment on the expected size (\ref{sec:Size}), summarise the implementation of the algorithm for the generation of configuration matrices (\ref{sec:Generation}) and discuss the algorithm for the calculation of the Hodge numbers (\ref{sec:Algorithm_Hodge}). The subset of the dataset that we restricted ourselves to is presented in section \ref{sec:Properties_dataset}, together with its properties. Section \ref{sec:NN} introduces the architectures that we chose for the neural network analysis of the Hodge numbers. Our findings are described in section \ref{sec:Results}, where each subsection focuses on the performance of Machine Learning (ML) in the prediction of one of the invariants. Finally, section \ref{sec:Conclusions} consists of the conclusions and outlook. The appendix contains the proof of finiteness of CICY five-folds, and detailed examples illustrating the generation of the dataset and the Hodge numbers calculation.

\section{Constructing and Characterising Calabi-Yau Five-folds}
\label{sec:2}
In this section we provide a quick overview on complete intersection Calabi-Yau manifolds in products of projective spaces. For more details on Calabi-Yau spaces in general, the reader is referred to \cite{Nakahara:2003nw, Besse:1987pua, Gross2003CalabiYauMA}, while some relevant sources for the construction of CICY's are ~\cite{CANDELAS1988493, 1987CMaPh.108..291H, Hubsch:1992nu, https://doi.org/10.48550/arxiv.1804.08792}.

\subsection{The Construction}
\label{sec:The_construction}
As mentioned in the introduction, CICY's are defined as smooth submanifolds in the product of projective spaces. We can write such a product generically as $\mathbb{CP}^{n_1} \times \cdots \times \mathbb{CP}^{n_m}$, and it is usually referred as the \textit{ambient space}, which we denote as $\mathcal{A}$. In order to consistently define a non-trivial submanifold, we consider homogeneous polynomials in the ambient space. Let us label them as $p_{\alpha}$, with $\alpha= 1,...,K$. Each $p_{\alpha}$ depends on $m$ sets of homogeneous coordinates (one for each projective space in the ambient space). Then, in general, the zero locus of these polynomials defines a smooth submanifold of dimension $\mathrm{dim}(\mathcal{A}) - K$. It turns out that the holomorphic numbers of the spaces originating from this construction do not depend on the specific form of the homogeneous polynomials, but only on their degree. We therefore define the following notation: the degree of the polynomial $p_{\alpha}$ in the $r^{\mathrm{th}}$ set of coordinates, associated to the $r^{\mathrm{th}}$ projective space, is denoted as $q_{\alpha}^r$. All the relevant information concerning the submanifold can thus be arranged in a \textit{configuration matrix}:
\begin{align}
    (\mathbf{n} | \mathbf{q}) \equiv\left(\begin{array}{c|ccc}
n_1 & q_1^1 & \ldots & q_K^1 \\
\vdots & \vdots & \ddots & \vdots \\
n_m & q_1^m & \ldots & q_K^m
\end{array}\right),
\label{eq:Conf_Mat}
\end{align}
which is the object that is commonly used to label CICY's. Any space obtained according to the prescription above is compact and K\"ahler, by definition. In order to obtain a manifold in complex dimension five, the condition
\begin{align}
K = \sum_r^m n_r -5
\label{eq:K_and_n}
\end{align}
must be satisfied. Moreover, so far we have only achieved the "CI" in "CICY". In order to ensure the Calabi-Yau property (which is defined, together with its implications, in the next section), we also need to impose a constraint on the degrees of the polynomials:
\begin{align}
    \sum_{\alpha=1}^K q_\alpha^r=n_r+1 .
    \label{eq:CY_Condition}
\end{align}
We conclude this section by introducing an important point, which is discussed in more details in section \ref{sec:Generation}: more than one configuration matrix may correspond to the same manifold. For example, it is evident that permuting columns does not change the nature of the space described, since it is just a relabelling of the polynomials. The same goes for permuting rows, which just amounts to changing the order of the projective spaces in the ambient space. These two observations already show that there is a very large amount of redundancy in the description of CICY via configuration matrices. There are other sources of redundancy as well, which are also described in section \ref{sec:Generation}.

\subsection{Basic Properties, Index Theorem and Euler Number}
\label{sec:Basic_properties}
This section follows closely Appendix B of \cite{Haupt:2008nu}, which is itself very similar to the detailed construction of four-folds that can be found in \cite{Gray_2013}.\\
Let us consider a complex 5-dimensional K\"ahler manifold $\mathcal{M}$. If its canonical bundle is trivial, or equivalently if its global holonomy group is contained in $ \tx{SU}(5)$, then we say that $\mathcal{M}$ is Calabi-Yau five-fold. This also implies that its first Chern class vanishes, i.e., $c_1(\mathcal{M}) = 0$. The Hodge numbers $h^{r,s}(\mathcal{M})$ are the holomorphic invariants used to classify the above manifolds. $h^{r,s}$ is defined as the complex dimension of the Dolbeault cohomology group $H^{r,s}(\mathcal{M})$ of $\mathcal{M}$, namely the $r^{\mathrm{th}}$ sheaf cohomology of the sheaf of germs of holomorphic $s$-forms:
\begin{equation}
    h^{r,s}(\mathcal{M}) = \tx{dim}\,H^{r,s}(\mathcal{M}) = \tx{dim} \,H^s(\mathcal{M}, \Omega_{\mathcal{M}}^r).
\end{equation}
Let us summarise some mathematical facts that are useful for calculating the Hodge numbers of Calabi-Yau five-folds.
It follows from Bochner techniques that $h^{0,p}(\mathcal{M})=h^{p,0}(\mathcal{M})=0$ for $p=1,2,3,4$. The Calabi-Yau condition, namely the triviality of the canonical bundle, ensures that $h^{0,0}(\mathcal{M})=h^{0,5}(\mathcal{M})=h^{5,0}(\mathcal{M})=h^{5,5}(\mathcal{M})=1$ (where we also implicitly used the Poincar\'e duality). Finally, we have the following symmetries between Hodge numbers: $h^{p,q}(\mathcal{M}) = h^{q,p}(\mathcal{M})$, by conjugation, and $h^{p,q}(\mathcal{M})=h^{5-p,5-q}(\mathcal{M})$, by Serre duality. Overall, the Hodge diamond of a Calabi-Yau five-fold is of the form
\begin{align}
 \begin{array}{ccccccccccc}
           &          &          &          &          &     1   &          &          &          &          &            \\
           &          &          &          &    0    &          &   0     &         &          &           &            \\
           &          &          &    0    &          &h^{1,1}&        &   0    &          &          &             \\ 
           &          &   0    &           &h^{1,2}&          &h^{1,2}&     &  0      &         &              \\     
           &  0      &         &h^{1,3}&          &h^{2,2}&        &h^{1,3}&     &     0    &              \\ 
    1    &          &h^{1,4}&          &h^{2,3}&          &h^{2,3}&        &h^{1,4}&    &    1     \\
          &  0      &         &h^{1,3}&          &h^{2,2}&        &h^{1,3}&     &     0    &              \\ 
           &          &   0    &           &h^{1,2}&          &h^{1,2}&     &  0      &         &              \\     
           &          &          &    0    &          &h^{1,1}&        &   0    &          &          &             \\ 
           &          &          &          &    0    &          &   0     &         &          &           &            \\
           &          &          &          &          &     1   &          &          &          &          & .
 \end{array} 
 \label{hodgediamond2.2}
\end{align} 
Given the usual decomposition for the Betti numbers,
\begin{equation}
    b_p(\mathcal{M}) = \sum_{r+s = p}h^{r,s}(\mathcal{M}),
\end{equation}
and the Hodge diamond \eqref{hodgediamond2.2}, it follows that
\begin{align}
\begin{array}{lllllll}
b_0(\mathcal{M})&=&1&\quad&b_1(\mathcal{M})&=&0 \\
 b_2(\mathcal{M})&=&h^{1,1}(\mathcal{M})&\quad&b_3(\mathcal{M})&=&2h^{1,2}(\mathcal{M}) \\
 b_4(\mathcal{M})&=&2h^{1,3}(\mathcal{M})+h^{2,2}(\mathcal{M})&\quad&b_5(\mathcal{M})&=&2+2h^{1,4}(\mathcal{M})+2h^{2,3}(\mathcal{M}) \, ,
 \end{array}
 \label{Betti_and_Hodge}
\end{align} 
and $b_i(\mathcal{M})=b_{10-i}(\mathcal{M})$ for $i>5$, by Poincar\'e duality. This, in turn, implies that the Euler number $\eta (\mathcal{M})$ of $\mathcal{M}$ can be written as 
\begin{equation}\begin{split}
 \eta (\mathcal{M})& = \sum_{i=0}^{10}(-1)^ib_i(\mathcal{M})\\
 &=2h^{1,1}(\mathcal{M})-4h^{1,2}(\mathcal{M})
 +4h^{1,3}(\mathcal{M})
 +2h^{2,2}(\mathcal{M})-2h^{1,4}(\mathcal{M})-2h^{2,3}(\mathcal{M})\; .
 \end{split}
 \label{Euler0}
\end{equation}
As it is made clear in \cite{Haupt:2008nu}, there is one more condition that can be derived from $c_1({\mathcal{M}})=0$ for the Hodge numbers, which was derived (for the case of four-folds) in \cite{Sethi_1996}. First, let us state the general form of the Atiyah-Singer index theorem,
\begin{equation}
     \chi (\mathcal{M},\mathcal{V}) = \sum_{i=0}^{{\rm dim}(\mathcal{M})}(-1)^i{\rm dim}H^i(\mathcal{M},\mathcal{V})=\int_{\mathcal{M}}{\rm ch}(\mathcal{V})\wedge{\rm Td}(\mathcal{T}\mathcal{M})\; , \label{indtheorem}
\end{equation}
for a complex vector bundle $\mathcal{V}$ on $\mathcal{M}$ and with $\tx{ch}(\mathcal{V})$ being the Chern character of $\mathcal{V}$. We denote the Todd class of the complexified tangent bundle of $\mathcal{M}$ by $\tx{Td}(\mathcal{T}\mathcal{M})$. In our case, $\mathcal{V}$ is the bundle of holomorphic $q$-forms, $\mathcal{V}= \Omega^q_{\mathcal{M}}$ where $q=0,1,2,3$. The cohomology groups of these bundles can be related to the cohomology group of the complex manifold $\mathcal{M}$ as $H^i(\mathcal{M}, \Omega^q_{\mathcal{M}})\simeq H^{i,q}(\mathcal{M})$. The Chern class and character of the tangent bundle are computed as
\begin{equation}\label{cch}
    c(\mathcal{T}\mathcal{M}) = 1 + c_1(\mathcal{T}\mathcal{M}) + c_2(\mathcal{T}\mathcal{M}) + \cdots = \prod_i (1+x_i), \quad \tx{ch}(\mathcal{T}\mathcal{M}) = \sum_i e^{x_i}.
\end{equation}
In the above equations, we have implicitly used the splitting principle to write the bundle $\mathcal{TM}$ as as a direct sum of line bundles $\mathcal{L}_i$, with the $x_i$'s being their first Chern classes (see \cite{Hubsch:1992nu} or \cite{Nakahara:2003nw}).
Then, the integrand of the index theorem reads:
\begin{equation}
    {\rm ch}(\Omega^q_{\mathcal{M}})\wedge{\rm Td}(\mathcal{T}\mathcal{M}) = \sum_{i_1>i_2>\cdots > i_q}e^{-x_{i_1}}\cdots e^{-x_{i_q}} \prod_j \frac{x_{j}}{1-e^{-x_j}}.
\end{equation}
Expanding this expression and re-writing it in terms of Chern classes using \eqref{cch}, it follows from the index theorem~\eqref{indtheorem}:
\begin{align}
 \chi (\mathcal{M},\Omega^0_{\mathcal{M}})&= h^{0,0}-h^{1,0}+h^{2,0}-h^{3,0}+h^{4,0}-h^{5,0}
 \label{eq:Chi_0}
 \\&=\frac{1}{1440}\int_{\mathcal{M}}\left[-c_2c_1^3+c_1^2c_3-c_1c_4+3c_1c_2^2\right], \nonumber\\
 \chi (\mathcal{M},\Omega^1_{\mathcal{M}})&= h^{0,1}-h^{1,1}+h^{2,1}-h^{3,1}+h^{4,1}-h^{5,1}
 \label{eq:Chi_1}
 \\&=\frac{1}{480}\int_{\mathcal{M}}\left[-c_1^3c_2+c_1^2c_3-21c_1c_4+3c_1c_2^2-20c_5\right], \nonumber\\
 \chi (\mathcal{M},\Omega^2_{\mathcal{M}})&= h^{0,2}-h^{1,2}+h^{2,2}-h^{3,2}+h^{4,2}-h^{5,2}
 \label{eq:Chi_2}
 \\&=\frac{1}{720}\int_{\mathcal{M}}\left[-c_1^3c_2+c_1^2c_3-31c_1c_4+3c_1c_2^2+330 c_5\right], \nonumber\\
 \chi (\mathcal{M},\Omega^3_{\mathcal{M}})&= h^{0,3}-h^{1,3}+h^{2,3}-h^{3,3}+h^{4,3}-h^{5,3}
 \label{eq:Chi_3}
 \\
 &=-\frac{1}{720}\int_{\mathcal{M}}\left[-c_1^3c_2+c_1^2c_3-31c_1c_4+3c_1c_2^2+330 c_5\right] , \nonumber
\end{align}
where $c_i=c_i(\mathcal{T}\mathcal{M})$.
Now, as a consequence of $c_1=0$ and of the symmetries in \eqref{hodgediamond2.2}, we have that \eqref{eq:Chi_0} is automatically satisfied, and \eqref{eq:Chi_2}, \eqref{eq:Chi_3} describe the same equation. Subtracting \eqref{eq:Chi_1} from \eqref{eq:Chi_2}, and comparing the resulting equation to \eqref{Euler0}, yields
\begin{equation}
 \eta (\mathcal{M})=\int_{\mathcal{M}}c_5(\mathcal{TM}). \label{Euler}
\end{equation}
If we instead eliminate $c_5$, we find a constraint which depends solely on Hodge numbers:
\begin{equation}
 11 h^{1,1}(\mathcal{M})-10h^{1,2}(\mathcal{M})-h^{2,2}(\mathcal{M})+h^{2,3}(\mathcal{M})+10h^{1,3}(\mathcal{M})-11h^{1,4}(\mathcal{M})=0,
 \label{Hodgecons}
\end{equation}
reducing the number of Hodge numbers needed to characterise five-folds from six to five.

Everything that was derived so far applies to any Calabi-Yau manifold. As anticipated, we now specialise to Calabi-Yau manifolds constructed as complete intersections of polynomials living in a product of projective spaces. These manifolds do not exhaust all possible Calabi-Yau's, but allow a systematic construction of many examples. We now make use of all the definitions given in the previous section, with $\mathcal{A}$ denoting the ambient space and $\mathcal{M}$ being the complete intersection Calabi-Yau. We let $J_r$ be the Fubini-Study K\"ahler form associated with one of the factors in the ambient space, i.e. $\mathbb{CP}^{n_r}$ for some $r$. We assume it to be normalised as
\begin{align}
    \int_{\mathbb{CP}^{n_r}} J_r^{n_r} =1 .
    \label{Normalisation}
\end{align}
Then, $\eta(\mathcal{M})$ can be computed by using \eqref{Euler}, together with (again, from \cite{Haupt:2008nu}):
\begin{align}
c_2([\mathbf{n} \mid \mathbf{q}]) &=c_2^{r s} J_r J_s=\frac{1}{2} \sum_{r, s=1}^m\left[-\left(n_r+1\right) \delta^{r s}+\sum_{\alpha=1}^K q_\alpha^r q_\alpha^s\right] J_r J_s , \nonumber \\
c_3([\mathbf{n} \mid \mathbf{q}]) &=c_3^{r s t} J_r J_s J_t=\frac{1}{3} \sum_{r, s, t=1}^m\left[\left(n_r+1\right) \delta^{r s t}-\sum_{\alpha=1}^K q_\alpha^r q_\alpha^s q_\alpha^t\right] J_r J_s J_t, \nonumber \\
c_4([\mathbf{n} \mid \mathbf{q}]) &=c_4^{r s t u} J_r J_s J_t J_u=\frac{1}{4}\left[-\left(n_r+1\right) \delta^{r s t u}+\sum_{\alpha=1}^K q_\alpha^r q_\alpha^s q_\alpha^t q_\alpha^u+2 c_2^{r s} c_2^{t u}\right] J_r J_s J_t J_u , \nonumber \\
c_5([\mathbf{n} \mid \mathbf{q}]) &=c_5^{r_1 \ldots r_5} J_{r_1} \cdots J_{r_5} \nonumber \\
&=\frac{1}{5}\left[\left(n_r+1\right) \delta^{r_1 \ldots r_5}-\sum_{\alpha=1}^K q_\alpha^{r_1} \cdots q_\alpha^{r_5}+5 c_3^{\left(r_1 r_2 r_3\right.} c_2^{\left.r_4 r_5\right)}\right] J_{r_1} \cdots J_{r_5}.
\label{Chern_classes}
\end{align}
Computing the components of $c_n^{r_1...r_n}$ is just a matter of tedious, but straightforward, summations, which can be easily implemented on a computer. It should be noted that some non-zero components are associated to a combination of Kähler forms which vanishes. For instance, consider a product of $\mathbb{CP}^1$ and $\mathbb{CP}^5$. Then, in the expansion of $c_2$, we have a term of the form $c_2^{11} J_1 J_1$. In that case, despite $c_2^{11}$ might be non-zero, the term vanishes since $J_1^2$ is zero on $\mathbb{CP}^1$. Once $c_5$ is computed, and the vanishing terms in \eqref{Chern_classes} have been established, we need to integrate the non-zero terms according to \eqref{Euler} in order to obtain the Euler number. The integration is over $\mathcal{M}$, but we can translate it into an integral over the ambient space $\mathcal{A}$ using the relation
\begin{align}
    \int_{\mathcal{M}} w=\int_{\mathcal{A}} w \wedge \mu, \quad \mu=\bigwedge_{\alpha=1}^K\left(\sum_{r=1}^m q_\alpha^r J_r\right).
\end{align}
By considering the combinations that vanish and the normalisation \eqref{Normalisation}, the integral can be quickly evaluated.\footnote{Note that, for the case of having all $q_{\alpha}^r \neq 0$, remarkable simplifications occur, allowing to calculate all the Hodge numbers without the need of any spectral sequence (see \cite{Haupt:2008nu}). Unfortunately, there is a very small number of configuration matrices for CICY with all $q_{\alpha}^r \neq 0$.}\\
Let us now present the techniques used for the calculation of the cohomological numbers that we focus on in our investigations.

\subsection{Spectral sequences}
\label{sec:Spectral_sequences}
The calculation for the Hodge numbers involves the study of two exact sequences. The first one was already considered in the literature for the cases of three-folds and four-folds (see \cite{article_Green, Green:1987cr, Hubsch:1992nu, Gray_2013, Gray_2014}). On the other hand, the appearance of the second one is much more rare (an instance can be found in \cite{Dumachev2015CompleteIC}), but it is the key to tackle the more complicated problem of five-folds. We present them in order. 

\subsubsection{The Adjunction Sequence}
\label{sec:Adjunction}
This section is based on the lower-dimensional analogues of this work, mentioned above. We refer the reader to those sources for a detailed description, while only sketching the procedure here.  The analysis that we are about to present is sufficient to determine the full Hodge diamond for CICY three-folds and four-folds, but the same is not true for five-folds. The extra machinery necessary for the latter manifolds is described in the next subsection.

Let $\mathcal{T}_{\mathcal{A}}$ and $\mathcal{T}_{\mathcal{M}}$ be the holomorphic tangent bundles of $\mathcal{A}$ and $\mathcal{M}$, respectively. The normal bundle  to $\mathcal{M}$ is defined as the following quotient:  $\mathcal{N} = \mathcal{T_{\mathcal{A}}|_{\mathcal{M}} / \mathcal{T}_{\mathcal{M}}}$. Since $\mathcal{M}$ is embedded in $\mathcal{A}$ as the zero locus of some holomorphic section $\xi$, we obtain the short exact sequence:
\begin{align}
\left.\left.0 \rightarrow \mathcal{T}_{\mathcal{M}} \stackrel{i}{\rightarrow} \mathcal{T}_{\mathcal{A}}\right|_{\mathcal{M}} \stackrel{D \xi}{\longrightarrow} \mathcal{E}\right|_{\mathcal{M}} \rightarrow 0,
\label{eq:Adjunction_sequence}
\end{align}
where $\mathcal{E}$ is a holomorphic bundle over $\mathcal{A}$. This is called the adjunction sequence, and it is the key result for studying the cohomology of CICY's. It implies that $\mathcal{E}|_{\mathcal{M}} = \mathcal{N}$, which we can write as a sum of line bundles as
\begin{align}
    \mathcal{N} = \bigoplus_{a=1}^K \mathcal{O}_{\mathcal{M}}(\mathbf{q}_a) = \bigoplus_{a=1}^K \bigotimes_{r=1}^m\left(h_r\right)^{q_r^a},
\label{eq:Normal_bundle_as_sum_of_line_bundles}
\end{align}
where $\mathcal{O}_{\mathcal{A}}(k^r)$ is the line bundle with $c_1(\mathcal{O}_{\mathcal{M}}(k^r)) = k^r J_r$,  $\mathcal{O}_{\mathcal{M}}(k^r)$ is its restriction to $\mathcal{M}$ and $h_r$ denotes the hyperplane bundle over $\mathbb{CP}_r^{n_r}$.
The long cohomology sequence associated to \eqref{eq:Adjunction_sequence}, which is at the core of our computation of the Hodge numbers, reads:
\begin{align}
\begin{aligned}
 0 \xrightarrow[]{} 0 \cong H^0(\mathcal{M}, \mathcal{T}_{\mathcal{M}}) &\rightarrow   H^0\left(\mathcal{M}, \mathcal{T}_{\mathcal{A}|_\mathcal{M}}\right) \stackrel{}{\rightarrow} H^0(\mathcal{M}, \mathcal{E}) \rightarrow \\
 H^{4,1}(\mathcal{M}) \cong H^1(\mathcal{M}, \mathcal{T}_{\mathcal{M}}) &  \stackrel{}{\rightarrow} H^1\left(\mathcal{M}, \mathcal{T}_{\mathcal{A}|_\mathcal{M}}\right) \stackrel{}{\rightarrow} H^1(\mathcal{M}, \mathcal{E}) \rightarrow \\
H^{3,1}(\mathcal{M}) \cong H^2(\mathcal{M}, \mathcal{T}_{\mathcal{M}}) &  \stackrel{}{\rightarrow} H^2\left(\mathcal{M}, \mathcal{T}_{\mathcal{A}|_\mathcal{M}}\right) \stackrel{}{\rightarrow} H^2(\mathcal{M}, \mathcal{E}) \rightarrow \\
H^{2,1}(\mathcal{M}) \cong H^3(\mathcal{M}, \mathcal{T}_{\mathcal{M}})
&  \stackrel{}{\rightarrow} H^3\left(\mathcal{M}, \mathcal{T}_{\mathcal{A}|_\mathcal{M}}\right) \stackrel{}{\rightarrow} H^3(\mathcal{M}, \mathcal{E}) \rightarrow \\
H^{1,1}(\mathcal{M}) \cong H^4(\mathcal{M}, \mathcal{T}_{\mathcal{M}} )
&  \stackrel{}{\rightarrow} H^4\left(\mathcal{M}, \mathcal{T}_{\mathcal{A}|_\mathcal{M}}\right) \stackrel{}{\rightarrow} H^4(\mathcal{M}, \mathcal{E}) \rightarrow \\
0 \cong H^5(\mathcal{M}, \mathcal{T}_{\mathcal{M}} ) 
&  \stackrel{}{\rightarrow} H^5\left(\mathcal{M}, \mathcal{T}_{\mathcal{A}}|_\mathcal{M}\right) \stackrel{}{\rightarrow} H^5(\mathcal{M}, \mathcal{E}) \xrightarrow[]{} 0 . 
\label{eq:Long_adjunction_sequence}
\end{aligned}
\end{align}
As usual, Serre duality and Dolbeault theorem have been used to relate the cohomologies of the sequence to $H^{\bullet,1}(\mathcal{M})$.
In order to determine the Hodge numbers from $h^{1,1}$ to $h^{1,4}$, one first needs to calculate the cohomologies in the right two columns. 
The key object for the computation is the Koszul resolution:
\begin{align}
0 \longrightarrow \wedge^K \mathcal{E}^* {\longrightarrow} \cdots {\longrightarrow} \wedge^2 \mathcal{E}^* {\longrightarrow} \mathcal{E}^* {\longrightarrow} \mathcal{O}_X {\longrightarrow} \mathcal{O}_{\mathcal{M}} \longrightarrow 0,
\label{eq:Koszul_resolution}
\end{align}
In fact, by twisting the above sequence in precise ways we can determine the cohomologies valued in the tangent bundle (indirectly) and those valued in the normal bundle (directly). \\
Let us start from the former, for which we need yet another short exact sequence, the Euler sequence:
\begin{align}
    0 \xrightarrow{} \mathcal{O}^m_{\mathcal{M}} \xrightarrow{} \mathcal{R} \xrightarrow{} \mathcal{T}_{\mathcal{A}}|_{\mathcal{M}}  \xrightarrow{} 0.
\end{align}
The bundle $\mathcal{R}$ is defined as
\begin{align}
    \mathcal{R}=\bigoplus_{r=1}^m \mathcal{O}_{\mathcal{M}}\left(\mathbf{e}_i\right)^{\oplus\left(n_r+1\right)},
    \label{eq:Bundle_R}
\end{align}
where $\mathbf{e}_i$ are the unit vectors. The associated long sequence in cohomology reads
\begin{align}
\begin{aligned}
0 \xrightarrow[]{} H^0(\mathcal{M}, \mathcal{O}^m_{\mathcal{M}}) &\rightarrow   H^0\left(\mathcal{M}, \mathcal{R} \right) \stackrel{}{\rightarrow} H^0(\mathcal{M}, \mathcal{T}_{\mathcal{A}|\mathcal{M}}) \rightarrow \\
H^1(\mathcal{M}, \mathcal{O}^m_{\mathcal{M}}) &  \stackrel{}{\rightarrow} H^1\left(\mathcal{M}, \mathcal{R}\right) \stackrel{}{\rightarrow} H^1(\mathcal{M}, \mathcal{T}_{\mathcal{A}|\mathcal{M}}) \rightarrow \\
H^2(\mathcal{M}, \mathcal{O}^m_{\mathcal{M}})
&  \stackrel{}{\rightarrow} H^2\left(\mathcal{M}, \mathcal{R}\right) \stackrel{}{\rightarrow} H^2(\mathcal{M}, \mathcal{T}_{\mathcal{A}|\mathcal{M}}) \rightarrow \\
H^3(\mathcal{M}, \mathcal{O}^m_{\mathcal{M}})
&  \stackrel{}{\rightarrow} H^3\left(\mathcal{M}, \mathcal{R}\right) \stackrel{}{\rightarrow} H^3(\mathcal{M}, \mathcal{T}_{\mathcal{A}|\mathcal{M}}) \rightarrow \\
H^4(\mathcal{M}, \mathcal{O}^m_{\mathcal{M}}) 
&  \stackrel{}{\rightarrow} H^4\left(\mathcal{M}, \mathcal{R}\right) \stackrel{}{\rightarrow} H^4(\mathcal{M}, \mathcal{T}_{\mathcal{A}|\mathcal{M}}) \rightarrow
\\
H^5(\mathcal{M}, \mathcal{O}^m_{\mathcal{M}}) 
&  \stackrel{}{\rightarrow} H^5\left(\mathcal{M}, \mathcal{R}\right) \stackrel{}{\rightarrow} H^5(\mathcal{M}, \mathcal{T}_{\mathcal{A}|\mathcal{M}}) \xrightarrow[]{} 0
.
\end{aligned}
\end{align}
The first column on the left is easy to compute, since $H^0(\mathcal{M}, \mathcal{O}^m_{\mathcal{M}}) \cong \mathbb{C}^m$, $H^5(\mathcal{M}, \mathcal{O}^m_{\mathcal{M}}) \cong \mathbb{C}^m$ and all the other cohomologies are trivial. The central column can be determined by considering \eqref{eq:Bundle_R}.
Specifically, we can consider the twisted version of the Koszul resolution \eqref{eq:Koszul_resolution} for each of the basis vectors and the associated spectral sequence, to calculate the cohomology of each $\mathcal{O}_{\mathcal{M}}\left(\mathbf{e}_i\right)$. We note that spectral sequence in this case is trivial, i.e., the filtration converges immediately, for all CICY three-folds (see \cite{Green:1987cr}).  The cohomology valued in $\mathcal{R}$ is then obtained by summing all such contributions, raised to the appropriate power. This allows to determine the missing column, i.e., the cohomology valued in the ambient space tangent bundle restricted, which is needed in the adjunction long cohomology sequence \eqref{eq:Long_adjunction_sequence}. 

Regarding the normal bundle, the situation is simpler, without the need for introducing any other sequences. According to the decomposition \eqref{eq:Normal_bundle_as_sum_of_line_bundles}, we can find the cohomology valued in the co-normal bundle by summing the cohomology of the line bundles twisted by each of the constraints. To calculate all these contributions, we can again consider the twisted version of the Koszul resolution, and its spectral sequence.\footnote{To be precise, all the computations outlined above are performed with the dual bundles. We work with the dual version of \eqref{eq:Adjunction_sequence}, and we substitute $\mathbf{q}_{\alpha} \xrightarrow{} -\mathbf{q}_{\alpha}$ in all the steps.} \\
Usually, when considering the spectral sequences above, it is the case that they degenerate at the first page. However, this is not guaranteed in general, and the condition that forbids the degeneracy is referred as the \textit{obstruction} in \cite{Green:1987cr}. Before getting to it, let us point out that the procedure outlined above has already been studied in detail for the case of three-folds, and general formulae for the spectral sequences involved have been worked out. Closed expressions for the non-vanishing groups appearing in the spectral sequences, namely $E_1^{j,k}(\mathcal{T}_{\mathcal{A}})$ and $E_1^{j,k}(\mathcal{E})$, can be found in the appendix (again, see \cite{article_Green} and \cite{Green:1987cr} for more details). The obstruction is present whenever there is a pair of non-vanishing groups $E_1^{j,j}(\mathcal{E})$, $E_1^{j',j'-1}(\mathcal{E})$ such that $j \geq j'$. This implies the presence of non-trivial maps, which makes the filtration non-trivial. More examples of obstructions appear in the next section, and the specific approach that we chose to deal with this complication is discussed in section \ref{sec:Algorithm_Hodge}. \\
Finally, we recall that this machinery is not sufficient for determining the complete Hodge diamond of CICY five-folds. Hence, we need to consider yet another sequence, which brings us to the next subsection.

\subsubsection{The Symmetrised Adjunction Sequence}
\label{sec:Sym_adjunction}
The method just described allows us to find four Hodge numbers out of six (modulo the obstruction issue, which we discuss in detail). A naive counting would tell us that we have found the full diamond, since there are two extra constraints, namely \eqref{Euler0} and \eqref{Hodgecons}, coming from the Euler number and the index theorem that we discussed, respectively. However, $h^{2,2}$ and $h^{2,3}$ cannot be determined independently from those equations, since they appear in the same combination in both. Hence, the need for an additional computation, based on the symmetrisation of the adjunction sequence, which allows us to find the remaining Hodge numbers. An application of such a sequence can be found in \cite{Dumachev2015CompleteIC}, where only CICY's defined on a single projective space are studied. \\
The symmetric square of \eqref{eq:Adjunction_sequence} yields:
\begin{align}
    0 \xrightarrow[]{} \mathcal{T}_{\mathcal{M}}^{\,2} \xrightarrow[]{} \mathcal{T}_{\mathcal{A}}|_{\mathcal{M}}^{\,\,2} \xrightarrow[]{} \mathcal{T}_{\mathcal{A}}|_{\mathcal{M}} \otimes \mathcal{E}  \xrightarrow[]{} \mathrm{Sym}^2 \mathcal{E} \xrightarrow[]{} 0.
\end{align}
This sequence is not short, but it can be split into the following two short exact sequences:
\begin{align}
\begin{aligned}
    0 \xrightarrow[]{} \mathcal{T}_{\mathcal{M}}^{\,2} \xrightarrow[]{} \mathcal{T}_{\mathcal{A}}|_{\mathcal{M}}^{\,\,2} \xrightarrow[]{} \mathcal{K} \xrightarrow[]{} 0 , \\
    0 \xrightarrow[]{} \mathcal{K} \xrightarrow[]{} \mathcal{T}_{\mathcal{A}}|_{\mathcal{M}} \otimes \mathcal{E}  \xrightarrow[]{} \mathrm{Sym}^2 \mathcal{E} \xrightarrow[]{} 0.
\end{aligned}
\end{align}
The associated long sequences in cohomology read:
\begin{align}
\begin{aligned}
0 \xrightarrow[]{} 0 \cong H^0(\mathcal{M},\mathcal{T}_{\mathcal{M}}^{\,2}) &\rightarrow   H^0\left(\mathcal{M}, \mathcal{T}_{\mathcal{A}}|_{\mathcal{M}}^{\,\,2} \right) \stackrel{}{\rightarrow} H^0(\mathcal{M}, \mathcal{K}) \rightarrow  \\
H^{3,1}(\mathcal{M)} \cong
H^1(\mathcal{M},\mathcal{T}_{\mathcal{M}}^{\,2}) &  \stackrel{}{\rightarrow} H^1\left(\mathcal{M}, \mathcal{T}_{\mathcal{A}}|_{\mathcal{M}}^{\,\,2}\right) \stackrel{}{\rightarrow} H^1(\mathcal{M}, \mathcal{K}) \rightarrow \\
H^{2,3}(\mathcal{M)} \cong
H^2(\mathcal{M},\mathcal{T}_{\mathcal{M}}^{\,2})
&  \stackrel{}{\rightarrow} H^2\left(\mathcal{M}, \mathcal{T}_{\mathcal{A}}|_{\mathcal{M}}^{\,\,2}\right) \stackrel{}{\rightarrow} H^2(\mathcal{M}, \mathcal{K}) \rightarrow \\
H^{2,2}(\mathcal{M)} \cong
H^3(\mathcal{M},\mathcal{T}_{\mathcal{M}}^{\,2})
&  \stackrel{}{\rightarrow} H^3\left(\mathcal{M}, \mathcal{T}_{\mathcal{A}}|_{\mathcal{M}}^{\,\,2}\right) \stackrel{}{\rightarrow} H^3(\mathcal{M}, \mathcal{K}) \rightarrow \\
H^{2,1}(\mathcal{M)} \cong H^4(\mathcal{M},\mathcal{T}_{\mathcal{M}}^{\,2}) 
&  \stackrel{}{\rightarrow} H^4\left(\mathcal{M}, \mathcal{T}_{\mathcal{A}}|_{\mathcal{M}}^{\,\,2}\right) \stackrel{}{\rightarrow} H^4(\mathcal{M}, \mathcal{K}) \rightarrow
\\
0 \cong
H^5(\mathcal{M},\mathcal{T}_{\mathcal{M}}^{\,2}) 
&  \stackrel{}{\rightarrow} H^5\left(\mathcal{M}, \mathcal{T}_{\mathcal{A}}|_{\mathcal{M}}^{\,\,2}\right) \stackrel{}{\rightarrow} H^5(\mathcal{M}, \mathcal{K}) \xrightarrow[]{} 0
,
\label{eq:Symmetrised_cohom_sequence_1}
\end{aligned}
\end{align}
and
\begin{align}
\begin{aligned}
0 \xrightarrow[]{} H^0(\mathcal{M},  \mathcal{K}) &\rightarrow   H^0\left(\mathcal{M},  \mathcal{T}_{\mathcal{A}}|_{\mathcal{M}} \otimes \mathcal{E} \right) \stackrel{}{\rightarrow} H^0(\mathcal{M}, \mathrm{Sym}^2 \mathcal{E}) \rightarrow \\
H^1(\mathcal{M},  \mathcal{K}) &  \stackrel{}{\rightarrow} H^1\left(\mathcal{M},  \mathcal{T}_{\mathcal{A}}|_{\mathcal{M}} \otimes \mathcal{E}\right) \stackrel{}{\rightarrow} H^1(\mathcal{M}, \mathrm{Sym}^2 \mathcal{E}) \rightarrow \\
H^2(\mathcal{M},  \mathcal{K})
&  \stackrel{}{\rightarrow} H^2\left(\mathcal{M},  \mathcal{T}_{\mathcal{A}}|_{\mathcal{M}} \otimes \mathcal{E}\right) \stackrel{}{\rightarrow} H^2(\mathcal{M}, \mathrm{Sym}^2 \mathcal{E}) \rightarrow \\
H^3(\mathcal{M},  \mathcal{K})
&  \stackrel{}{\rightarrow} H^3\left(\mathcal{M},  \mathcal{T}_{\mathcal{A}}|_{\mathcal{M}} \otimes \mathcal{E}\right) \stackrel{}{\rightarrow} H^3(\mathcal{M}, \mathrm{Sym}^2 \mathcal{E}) \rightarrow \\
H^4(\mathcal{M},  \mathcal{K}) 
&  \stackrel{}{\rightarrow} H^4\left(\mathcal{M},  \mathcal{T}_{\mathcal{A}}|_{\mathcal{M}} \otimes \mathcal{E}\right) \stackrel{}{\rightarrow} H^4(\mathcal{M}, \mathrm{Sym}^2 \mathcal{E}) \rightarrow
\\
H^5(\mathcal{M},  \mathcal{K}) 
&  \stackrel{}{\rightarrow} H^5\left(\mathcal{M},  \mathcal{T}_{\mathcal{A}}|_{\mathcal{M}} \otimes \mathcal{E}\right) \stackrel{}{\rightarrow} H^5(\mathcal{M}, \mathrm{Sym}^2 \mathcal{E}) \xrightarrow[]{} 0
,
\label{eq:Symmetrised_cohom_sequence_2}
\end{aligned}
\end{align}
respectively. In the first sequence, we have related the cohomology of the tangent space squared to the Hodge numbers of the Calabi-Yau, again by using Serre duality, Dolbeault theorem and the symmetries of the Hodge diamond. By straightforward generalisations of the techniques used in the previous section, we can compute $H^{\bullet}\left(\mathcal{M}, \mathcal{T}_{\mathcal{A}}|_{\mathcal{M}}^{\,\,2}\right)$, $H^{\bullet}\left(\mathcal{M},  \mathcal{T}_{\mathcal{A}}|_{\mathcal{M}} \otimes \mathcal{E}\right)$ and $H^{\bullet}(\mathcal{M}, \mathrm{Sym}^2 \mathcal{E})$. Let us start from the last one. The cohomology can be found by summing the contributions coming from the line bundle twisted by every possible pair of constraints. Each of those is obtained by considering the associated version of the Koszul resolution. Analogously, $H^{\bullet}\left(\mathcal{M}, \mathcal{T}_{\mathcal{A}}|_{\mathcal{M}}^{\,\,2}\right)$ is determined by summing the cohomology of the line bundles twisted by each pair of unit vectors (see \eqref{eq:Bundle_R}). Again, all these terms are found by considering the appropriate twisted Koszul resolution. Finally, $H^{\bullet}\left(\mathcal{M},  \mathcal{T}_{\mathcal{A}}|_{\mathcal{M}} \otimes \mathcal{E}\right)$ is calculated by twisting $\mathcal{T}_{\mathcal{A}}|_{\mathcal{M}}$ by each of the constraints, and summing their cohomologies. \\
These three results can be plugged in \eqref{eq:Symmetrised_cohom_sequence_1} and \eqref{eq:Symmetrised_cohom_sequence_2}, from which the unknown dimensions of $H^{2,2}$ and $H^{2,3}$ can be determined, most of the times. We end this section by briefly discussing when this procedure might not be sufficient for determining the desired Hodge numbers. The first immediate observation is that, under the assumption that $H^{\bullet}\left(\mathcal{M}, \mathcal{T}_{\mathcal{A}}|_{\mathcal{M}}^{\,\,2}\right)$, $H^{\bullet}\left(\mathcal{M},  \mathcal{T}_{\mathcal{A}}|_{\mathcal{M}} \otimes \mathcal{E}\right)$ and $H^{\bullet}(\mathcal{M}, \mathrm{Sym}^2 \mathcal{E})$ are known, there might be too many non-zero entries in \eqref{eq:Symmetrised_cohom_sequence_1} and \eqref{eq:Symmetrised_cohom_sequence_2} to find the Hodge numbers uniquely. However, we find that this is not the case, while the real obstacle that we find is the assumption above. In fact, the computation of those three cohomologies involve the study of filtrations which are not always trivial, i.e. there might be an \textit{obstruction}, as mentioned previously. For such cases, the presence of additional maps introduces extra variables, which complicates the problem. In section \ref{sec:Algorithm_Hodge} we discuss how we tackled this issue when implementing the above procedure algorithmically.
 
\section{Building a Dataset}
\label{sec:Building_dataset}
Given a method for computing the Hodge numbers from a given configuration matrix, the next natural step is to construct a list of many such configuration matrices. The only subtle point in this procedure is to take redundancy into account. Specifically, a number of configuration matrices might describe the same space, in which case only one representative should be picked\footnote{One might argue that, for the purpose of machine learning, including redundancies is actually beneficial for the neural network since it increases the size of the dataset. This is obviously true, but since the CICY five-folds dataset has never been fully constructed before, we need to perform these checks in order to achieve a good understanding of it.}. As described in \cite{Gray_2013}, the first check is whether two matrices just differ by a permutation of rows and columns. There are also other known techniques to identify equivalent configurations (again, see \cite{Gray_2013}) which need to be implemented in order to build a dataset free of redundancies.

In this section, we comment on the global properties of the dataset that can be inferred a priori, we describe the process of generating configuration matrices and we outline the implementation of the Hodge numbers calculation.

\subsection{The Size}
\label{sec:Size}
Finiteness of CICY three-folds has been proven in \cite{Green:1986ck}, where the maximum size of a configuration matrices for such spaces is derived. By similar arguments, one can show that four-folds are also finite in number (see \cite{Gray_2013}), and clearly the same holds for five-folds. In fact, we find that configuration matrices for five-folds can be at most of size $25 \times 30$, i.e., $25$ projective space factors and $30$ constraints over them. The proof of this can be found in the appendix. However, the actual size of the dataset is a whole different story. The three-folds list consists of $7890$ spaces, while the four-folds one has $921497$ CICY's. Even though not all equivalences have been removed from those datasets, it is expected that the effective number of inequivalent configurations is of the same order of magnitude \cite{Gray_2013}, \cite{Anderson_2008}. Now, by considering that there are just a few two-folds, we can estimate the expected size of our dataset to be in the order of $10^8$. Even with current computational power, constructing the whole dataset would be a very challenging task, and thus we will be concerned with a subset of it, as described in section \ref{sec:Properties_dataset}.

\subsection{Generation and Redundancies}
\label{sec:Generation}
As mentioned in section \ref{sec:The_construction}, there might be more than one configuration matrix describing the same manifold. There are different reasons for this. The most intuitive one is that exchanging rows and columns corresponds to a simple relabelling of the projective spaces and constraints, respectively. Hence, it does not affect the construction, leading to the same Calabi-Yau. This insensitivity to permutations accounts for the largest part of the redundancy in the dataset, and we will shortly outline how it has been removed in our case. However, there are (at least) other two sources of equivalence, which we did not consider in this work. According to \cite{Gray_2013}, they consist of matrices related by ineffective splitting and those related by accidental identities. We refer the reader to the discussion therein for further details, and we just quickly comment on the nature of these three redundancies.

Row/column permutations relate matrices of the same dimensions and, as we mentioned, they constitute the vast majority of the equivalence class. On the other hand, the other two redundancies involve pair of matrices with different dimensions. This feature allows to trade small matrices for larger ones, with more zeroes, whose spectral sequence calculations simplify. We return to this point at the end of next section, and we now focus on our algorithm for removing permutation matrices.

Given a maximum size for the configuration matrix, the algorithm creates the possible choices of $\mathbf{n}$ such that the above bound is satisfied. The number of rows is simply given by the number of projective spaces, i.e., the length of $\mathbf{n}$, while the number of columns is immediately found according to \eqref{eq:K_and_n}. The $\mathbf{n}$'s found this way are arranged in lexicographic order, which is the zeroth way of avoiding redundancies. Then, for each element of the above list, there might be many configuration matrices satisfying the Calabi-Yau condition \eqref{eq:CY_Condition}, hence defining a CICY. Since such condition should be satisfied row by row, the algorithm finds the allowed combinations of $q_{\alpha}^r$ for each $r$. These sets of rows are again arranged in lexicographic order. To obtain all the possible configuration matrices out of these sets, we take the Cartesian product of them. Clearly, when $n_r = n_{r'}$ ($r \neq r'$), one should avoid repetitions when taking the product of the two sets, and the same is true if more than two $n_r$'s are equal. This is also implemented in the algorithm to avoid redundant calculations. The configuration matrices thus obtained all share the same column $\mathbf{n}$, i.e., they have the same ambient space, by construction. However, because of the way they were constructed, no matrices in the list can be related by permutations of rows and columns. For the next step, we choose one of such matrices, and we generate (possibly) new CICY configurations by doing permutations within each row. The strategy is the same as above: we create a permutation set for each row, and then take the Cartesian products of these sets, ignoring repetitions for identical rows. Now, this list might contain matrices that are related by permutations of rows and columns. To check it, we employ the efficient algorithm described in \cite{Gray_2013} (and using brute force when the eigenvalues are degenerate). We now repeat the last procedure, where we permute elements within each row and check equivalence, for all the matrices that share the same $\mathbf{n}$. And, finally, we move on to the next $\mathbf{n}$ in the list. \\
The above procedure can be summarised schematically as follows. Given some bounds for the dimensions of the configuration matrix, we perform these steps, in order: 
\begin{enumerate}
    \item Find the the $\mathbf{n}$'s (ambient spaces) that satisfy the bounds, and order them lexicographically.
    \item Pick an $\mathbf{n}$, and build all the possible matrices out of the lexicographically ordered rows satisfying the Calabi-Yau condition.
    \item Pick one matrix of the above, and build all the matrices obtained by permuting each row independently.
    \item Identify and remove matrices that are related by permutations of rows and columns.
    \item Repeat for all the matrices obtained in 2, and for all $\mathbf{n}$'s obtained in 1.
\end{enumerate}
This procedure is best illustrated with an example, which can be found in the appendix. \\
To conclude this section, we note that building the whole dataset (up to matrices with dimensions $25 \times 30$) with the algorithm just described is hopeless with the computational resources at our disposal. In fact, this was already noted in \cite{Gray_2013}, when constructing all CICY four-folds. They employed a version of the above algorithm to produce only matrices that correspond to spaces without $\mathbb{CP}^1$ factors. The rest of the dataset is obtained by performing effective splittings to produce inequivalent configurations, until it is no longer possible to do so. This is a natural extension of our work, as we discuss in the next section.

\subsection{Algorithm for the Hodge Numbers}
\label{sec:Algorithm_Hodge}
We now outline our implementation of the procedure described in section \ref{sec:Spectral_sequences}.
Regarding the (standard) adjunction sequence, we implement the calculation using the formulae presented in \cite{article_Green}, which can be found in the appendix. They describe the first page of the spectral sequences associated to the ambient space tangent bundle and the normal bundle, which allow to determine the cohomologies that appear in \eqref{eq:Long_adjunction_sequence}. As we discussed in section \ref{sec:Spectral_sequences}, it is not always the case that the spectral sequence degenerates at the first page (see the \textit{obstruction} mentioned above). In fact, when there are maps connecting two non-zero entries we need to introduce a new variable for each of those, representing the kernel of the map. We refer to all of the extra variables that might appear in the computation, but are not the Hodge numbers, as \textit{supplementary variables}. Hence, in general, one obtains more unknowns than the four Hodge numbers appearing in \eqref{eq:Long_adjunction_sequence}, with a number of equations that depend on how many zeroes appear in the sequence and on their positions. For each configuration matrix, our algorithm determines such a system of equations and proceeds to find the unique solution whenever the system is well-posed. When it is not well-posed, we
still proceed to find a number of solutions, because it might be the case that the system is underconstrained only in the supplementary variables.\footnote{Note that, by construction, the system cannot turn out to be unsolvable.} Solutions are found within a region bounded by the inequalities that follow from the cohomology sequence. Specifically, for any exact sequence $0 \xrightarrow[]{} A_1 \xrightarrow[]{} A_2 \xrightarrow[]{} \cdots \xrightarrow[]{} 0$, we have that the dimensions $a_i = \mathrm{dim}(A_i)$ satisfy
\begin{align}
    (-1)^{I} \left(\sum_{i = 1}^{i = I} (-1)^i a_i \right) \geq 0 \, .
\end{align}
Applying this to the long cohomology sequence \eqref{eq:Long_adjunction_sequence} yields bounds on the variables involved, ensuring finiteness of solutions. To solve the system of linear equations, with the above bounds, we employ MiniZinc. Once the software has found a number of solutions, we can identify three possible scenarios:
\begin{itemize}
    \item The solution is unique.
    \item The solution is not unique, in which case there are two further sub-cases.
    \begin{itemize}
        \item The supplementary variables decouple, and the solution is unique in the Hodge numbers.
        \item There are solutions with different sets of Hodge numbers.
    \end{itemize}
\end{itemize}
In the first case and in the first sub-case, the Hodge numbers are successfully determined. On the other hand, for the second sub-case, our approach fails to provide a definite numerical answer. All the configuration matrices for which we were not able to determine the full Hodge diamond fall into this category. Now, clearly, this does not mean that their Hodge numbers cannot be calculated. The way to proceed is to consider the ineffective splitting equivalence. This equivalence allows to trade small matrices for large matrices, which describe the same manifold, but provide significant simplification in calculating the cohomological properties. The ineffective splitting is ubiquitous in the literature of CICY (see \cite{Green:1987cr, Gray_2013, Anderson_2017}, for instance), and it is a very powerful tool. Preliminary investigations show that it is as useful for five-folds as it was for the lower-dimensional analogues. However, we do not apply it to our sub-dataset, in the light of the considerations made at the end of the previous section. It is more natural to implement the ineffective splitting, together with the effective splitting algorithm, in a unified effort to produce and classify a larger set of CICY five-folds. Hence, we leave this for future work.

\section{Properties of the \texorpdfstring{$\mathbf{4 \times 4}$ Dataset}{Lg}}
\label{sec:Properties_dataset}
\begin{figure}[H]\label{fig:h11}
  \includegraphics[scale=0.33]{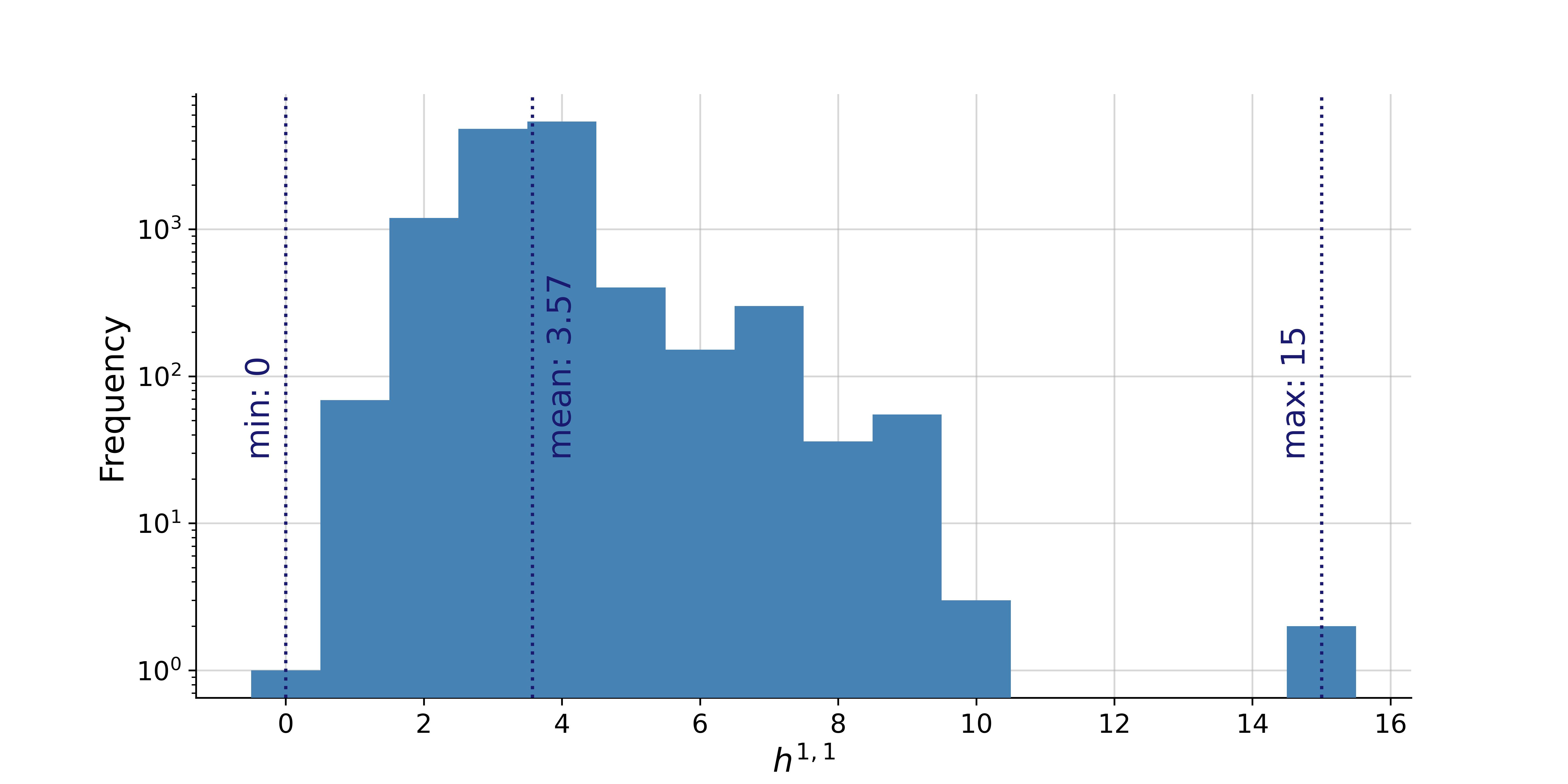}
  \includegraphics[scale=0.33]{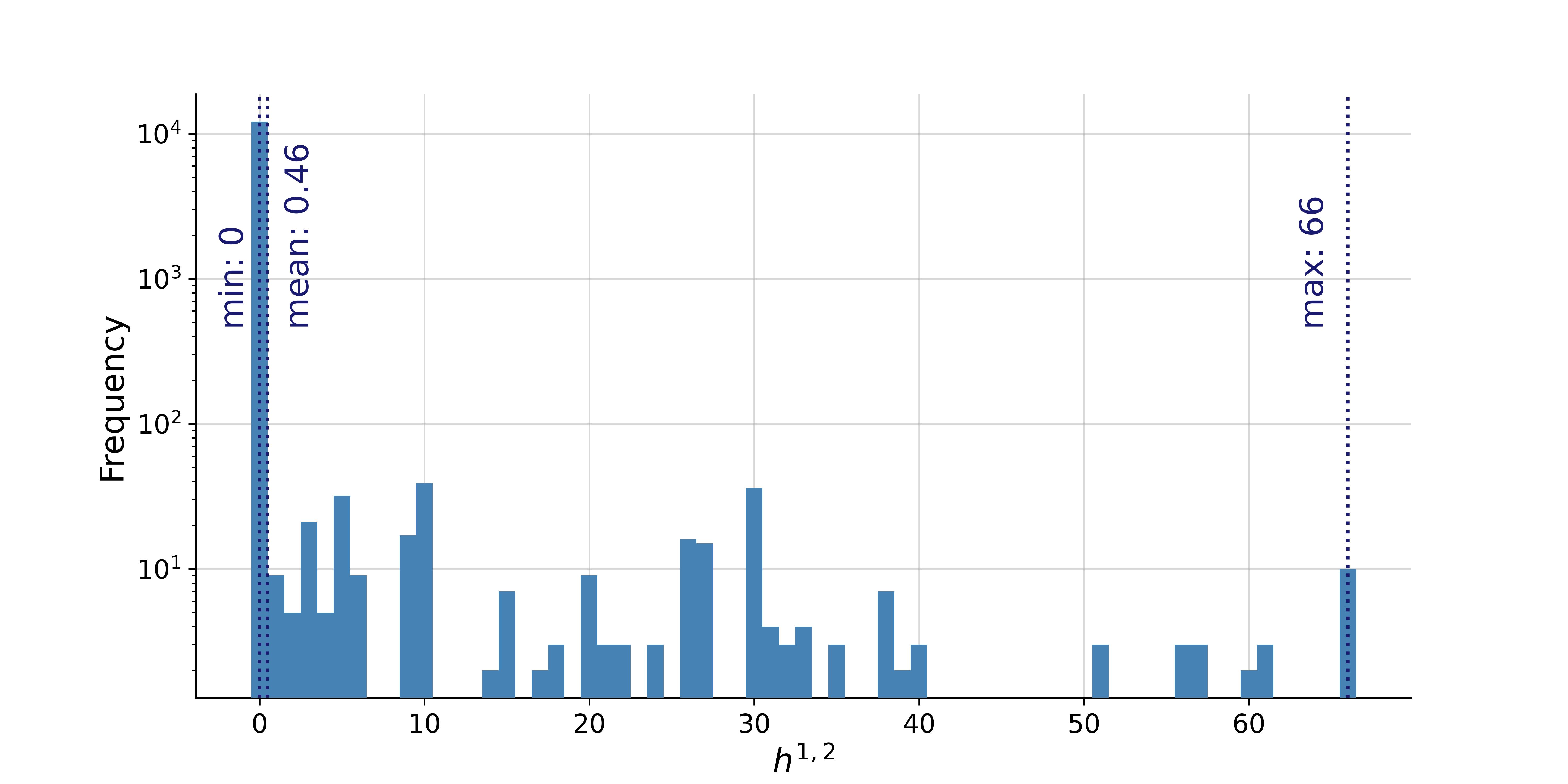}
  \includegraphics[scale=0.33]{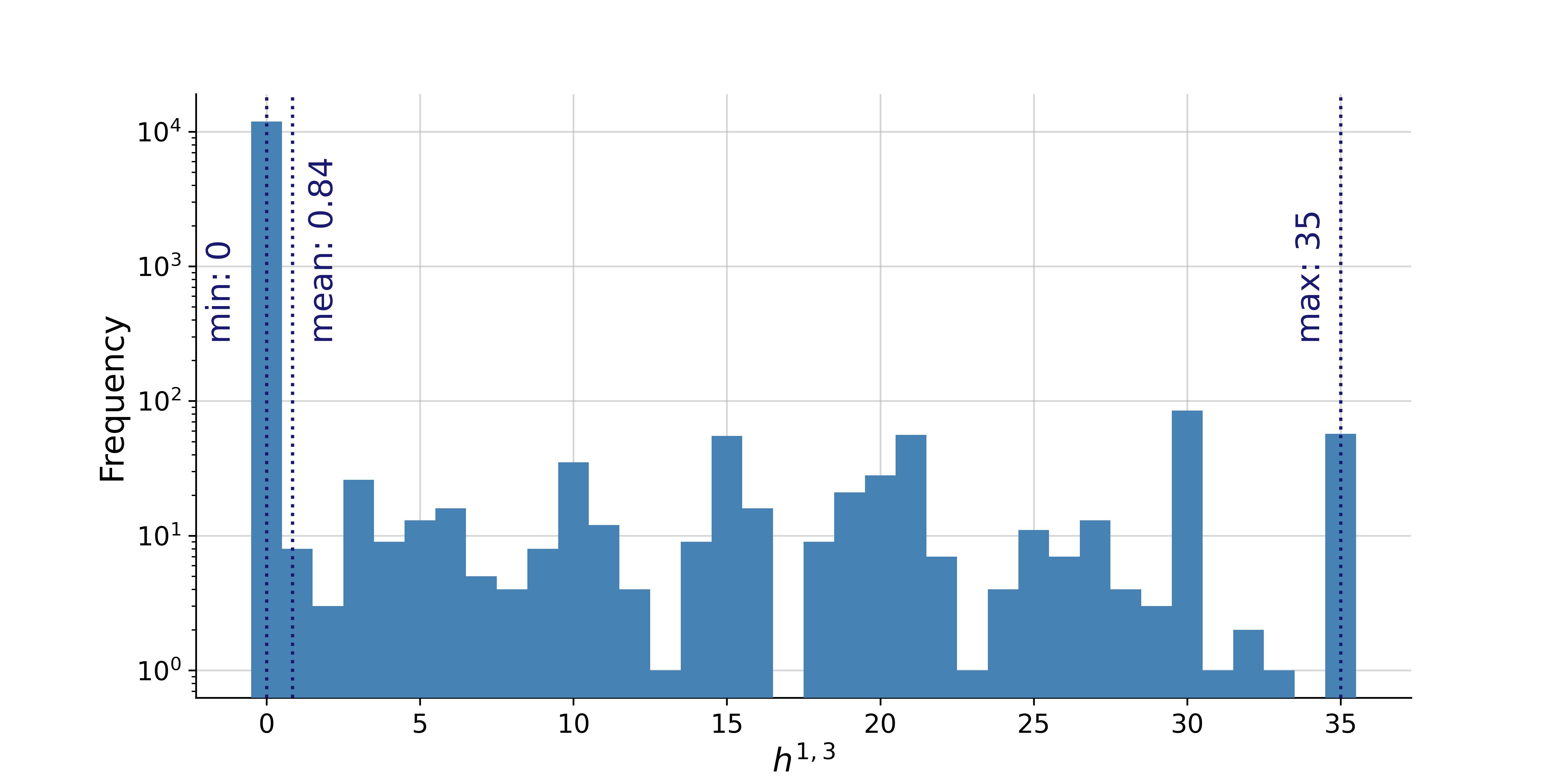}
  \includegraphics[scale=0.33]{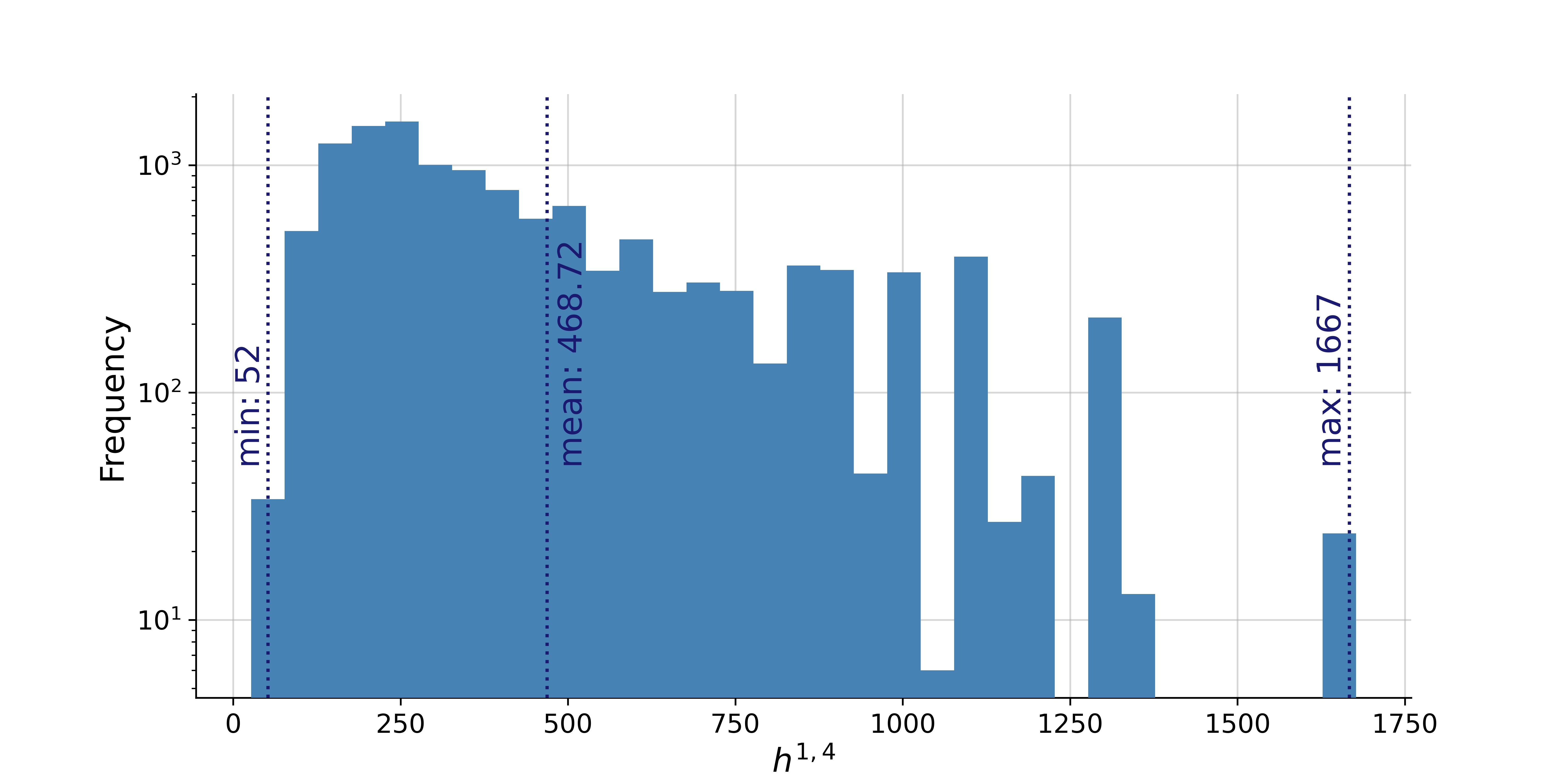}
  \includegraphics[scale=0.33]{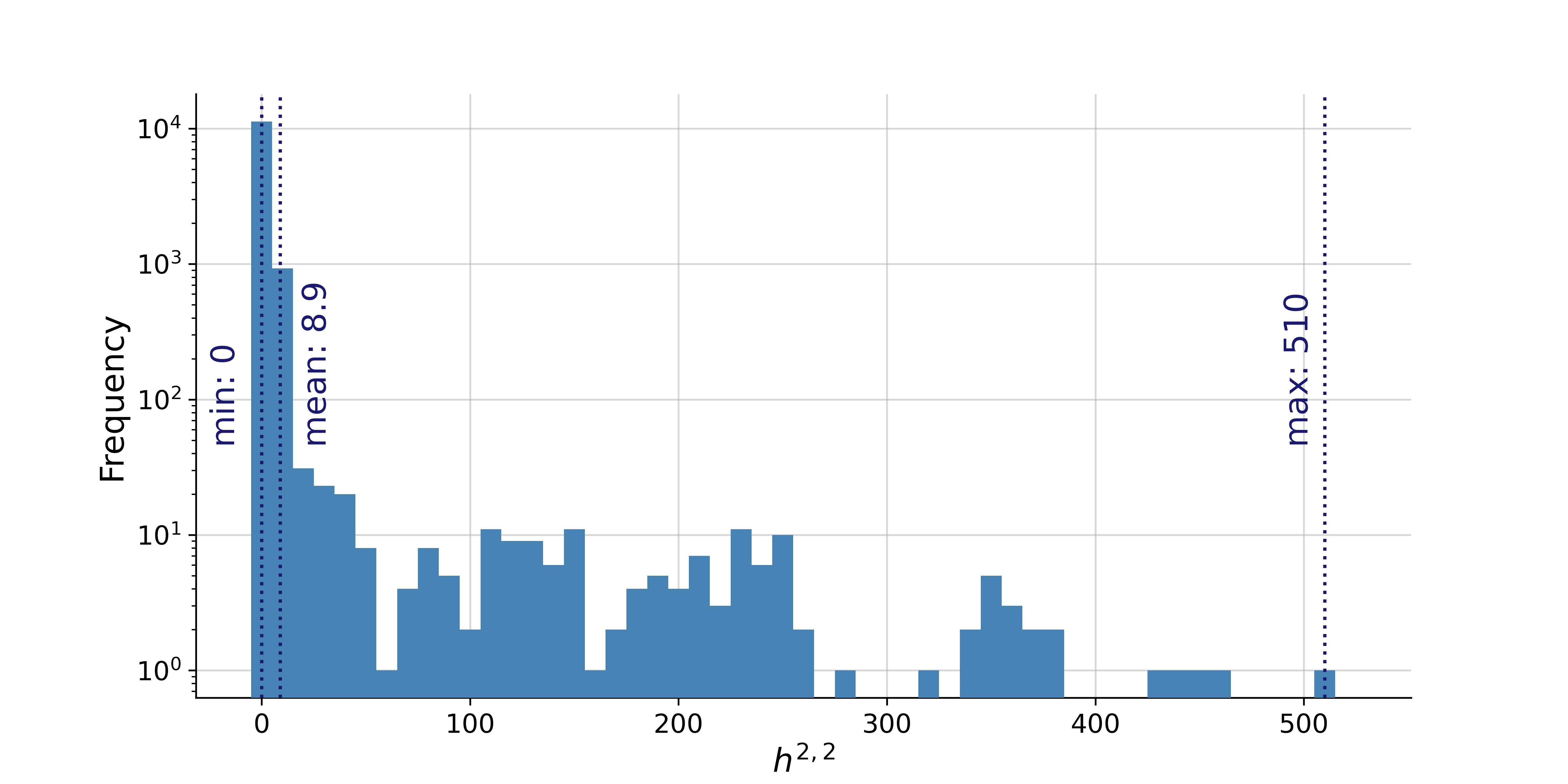}
  \includegraphics[scale=0.33]{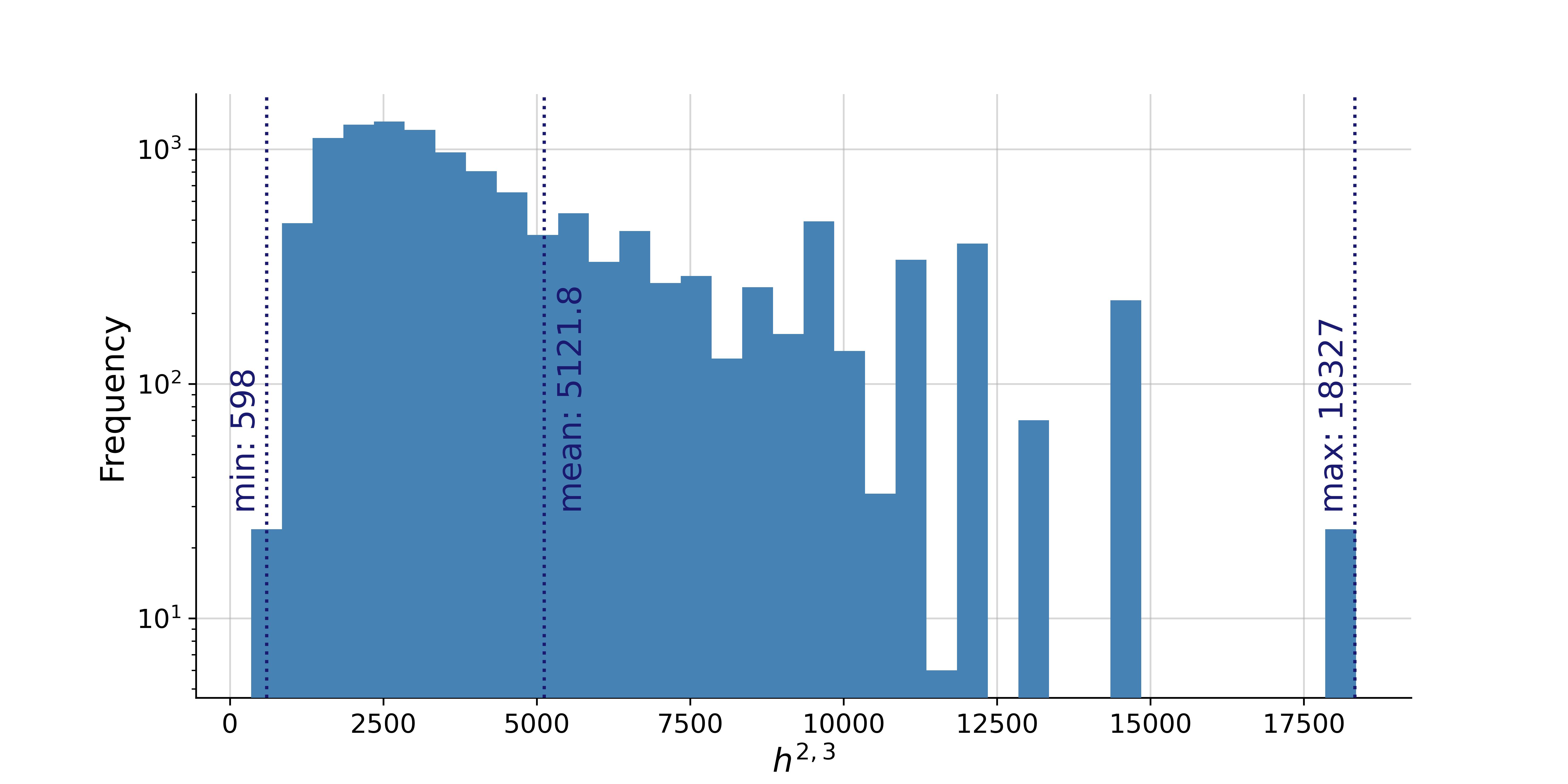}
  \begin{center}
    \includegraphics[scale=0.33]{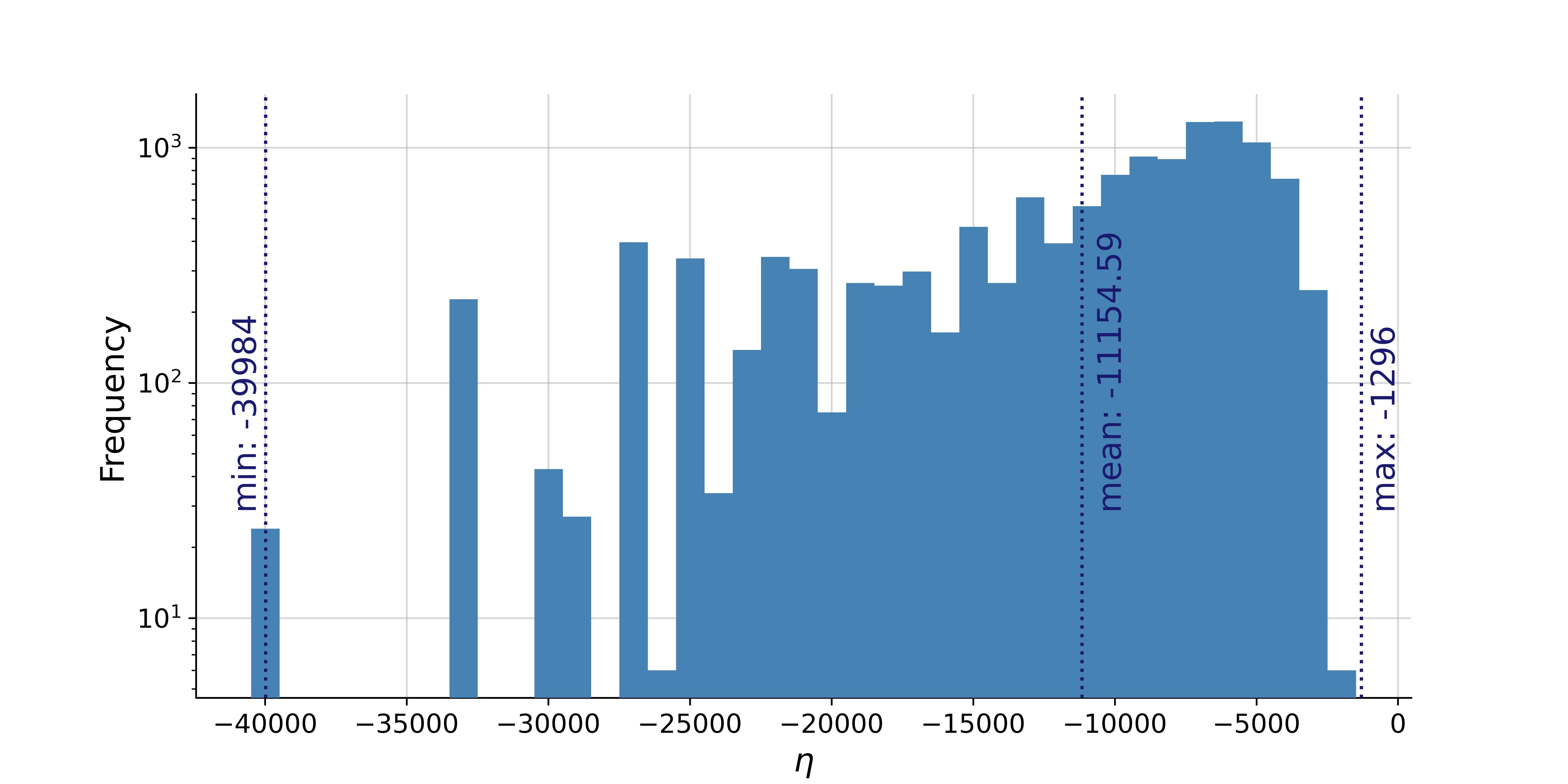}
  \end{center}
  \caption{These plots show the distribution of the invariants that we computed. Since $h^{1,1}$ has the smallest range, it is the most suitable piece of data in the machine learning context. Conversely, we can anticipate that the wide ranges of $h^{1,4}$, $h^{2,2}$ and $h^{2,3}$ make it very hard for the neural network to predict those numbers exactly. Finally, it is interesting to note how $h^{1,2}$ and $h^{1,3}$ vanish in the vast majority of cases, resulting in a very skewed sampling.}
  \label{fig:Histograms}
\end{figure}

\begin{figure}[H]\label{fig:Comparison_of_dim}
\centering
\includegraphics[scale=0.48]{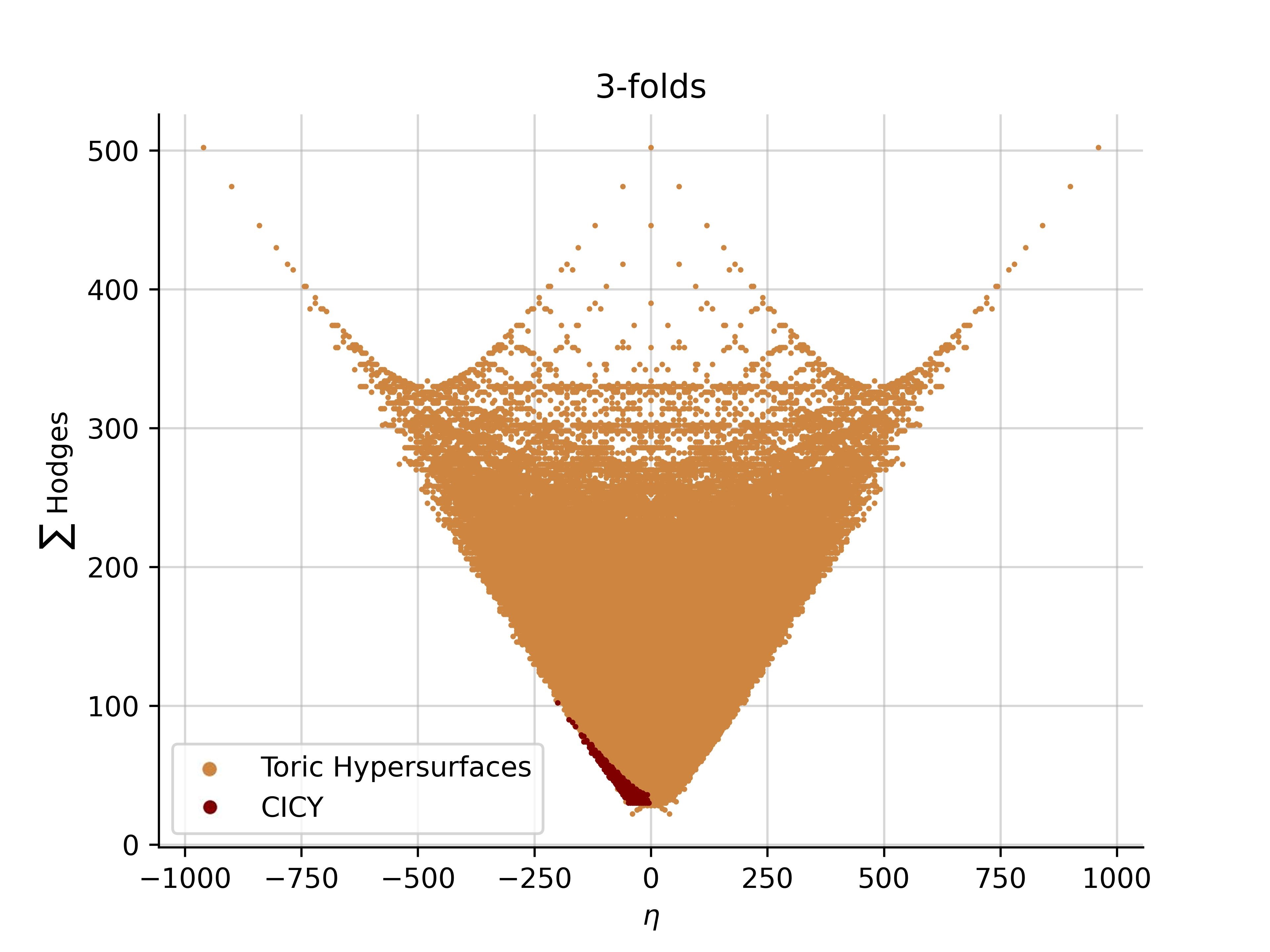}
\includegraphics[scale=0.48]{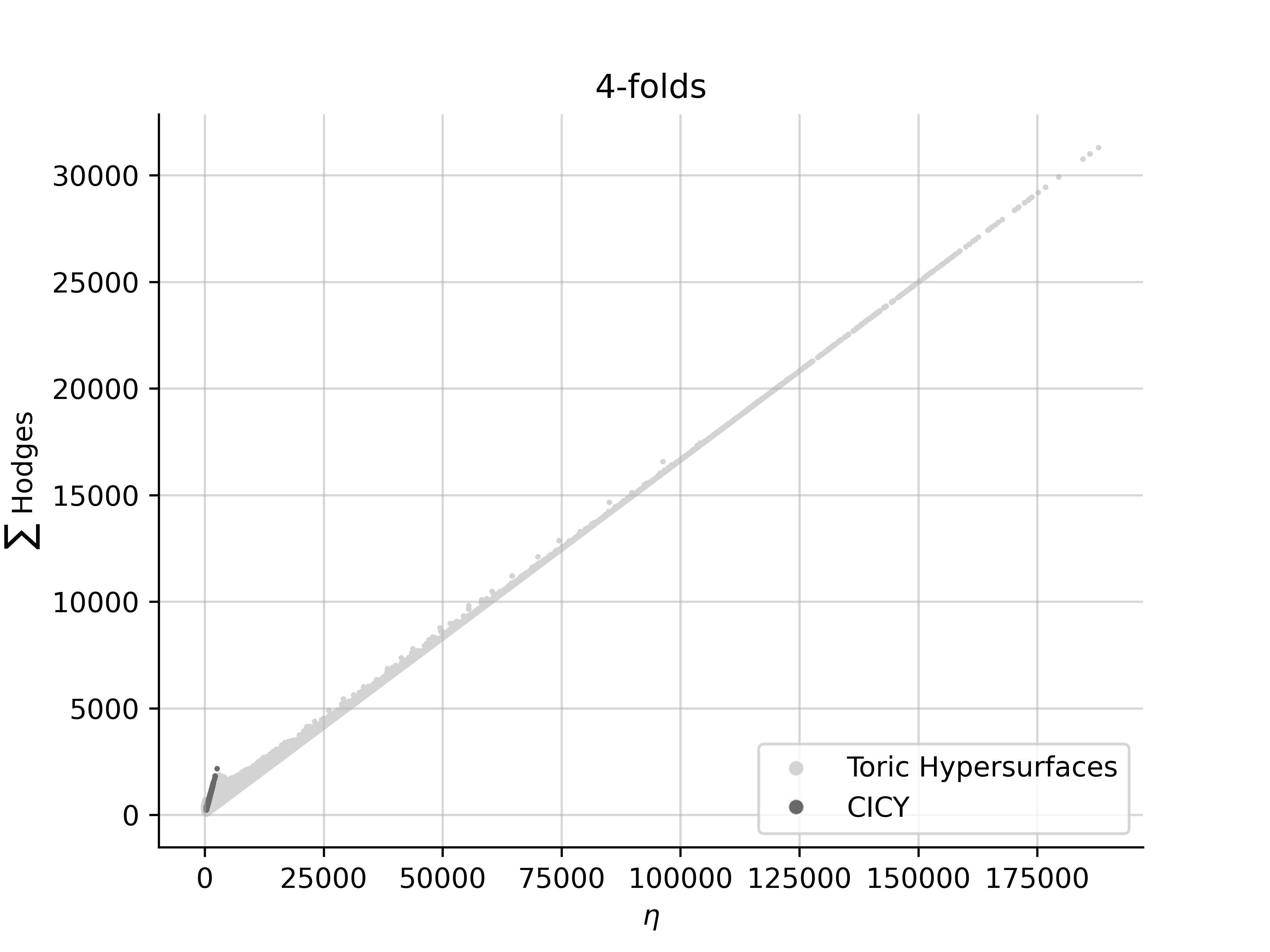}
\includegraphics[scale=0.14]{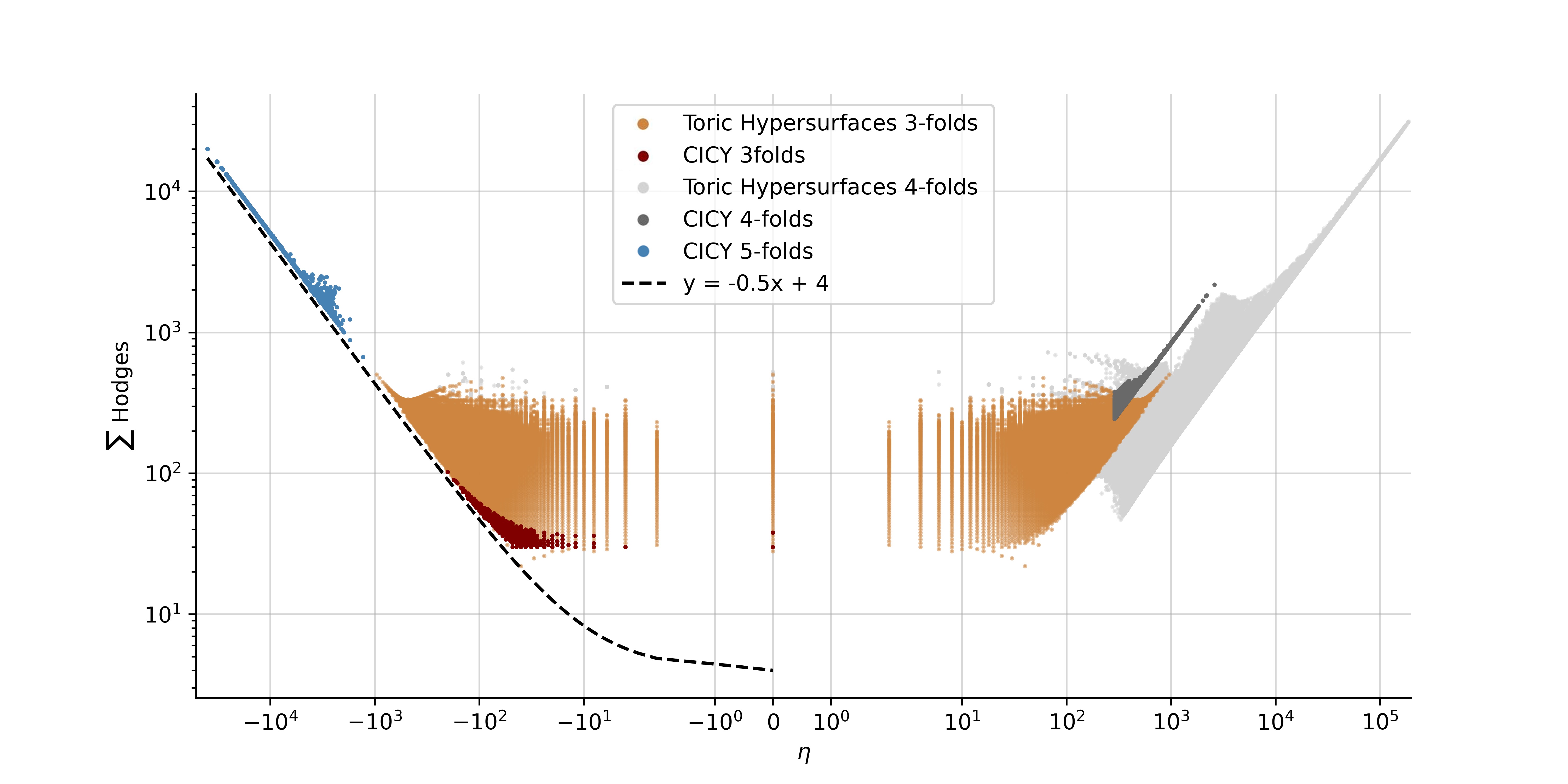}
  \caption{The plots represent Hodge data of Calabi-Yau spaces of different dimensions, obtained via different constructions. CICY three-folds and CICY four-folds data were taken from \cite{3F_data} and \cite{4F_data}, respectively. The toric hypersurfaces data can be found at \cite{KSdata}.}
  \label{fig:2}
\end{figure}

In this section, we present and comment on the properties of the dataset constructed according to the techniques just described. It is available \href{https://www.dropbox.com/scl/fo/z7ii5idt6qxu36e0b8azq/h?rlkey=0qfhx3tykytduobpld510gsfy&dl=0}{\textcolor{blue}{here}}. \\
Due to the expected astronomical size (see section \ref{sec:Size}), we chose to restrict ourselves to matrices with dimensions up to $4 \times 4$. This yielded $27068$ configuration matrices, out of which $3909$ are products. We believe that this number is large enough to exhibit some general properties, and it is definitely large enough to allow an efficient application of machine learning techniques. In fact, very promising results were obtained in \cite{He:2020lbz}, when restricting to configuration matrices with size up to $4 \times 4$, for CICY four-folds. We have been able to compute the full Hodge diamond for $12433$ matrices, i.e. $53.7 \%$ of the non-product matrices. We computed the Euler number for all the configuration matrices in the dataset.\\
Let us now present some general properties of our results. The distributions of the Hodge numbers and the Euler number are shown in figure \ref{fig:Histograms}, where we also plot mean, upper bound and lower bound.
\graphicspath{ {./images/} }
The number of values that each invariant can take, with configuration matrices up to $4 \times 4$ and representing non-product spaces, is given in table \ref{tab:my_label} below.
\begin{table}[H]
    \centering
    \begin{tabular}{|c|c|c|c|c|c|c|c|c|}
    \cline {2 - 9}
      \multicolumn{1}{c|}{} & $h^{1,1}$  & $h^{1,2}$ & $h^{1,3}$ & $h^{1,4}$  & $h^{2,2}$  & $h^{2,3}$ & $\eta$  & Full Diamond  \\
      \hline
      Number of values & 12 & 33 & 34 & 642 & 150 & 2103 & 632 & 2375 \\
        \hline
    \end{tabular}
    \caption{Total number of distinct values for the cohomological quantities of interest.}
    \label{tab:my_label}
\end{table}
%$h^{1,1}: 12$, $h^{1,2}: 33$, $h^{1,3}: 34$, $h^{1,4}: 642$, $h^{2,2}: 150$, $h^{2,3}: 2103$, $\eta:632$, $(\mathrm{Diamonds}): 2375$. 
Just for reference, there are $265$ different Hodge diamonds among all CICY three-folds, and $4417$ distinct Hodge sets in the CICY four-folds dataset. \\
It is also interesting to compare our work with previous results in lower-dimensional Calabi-Yau spaces. We do so in figure \ref{fig:2}, where a lot of information is encoded.
\graphicspath{ {./images/} }
We start by reproducing the (famous) scattered plot with Calabi-Yau three-folds constructed as toric hypersurfaces, in orange, and as complete intersections, in red. This can be found in \cite{Ashmore_2012}, for instance. By the notation $\sum \tx{Hodges}$, we mean the sum of the non-trivial Hodge numbers. For three-folds, they are $h^{1,1}$ and $h^{1,2}$, and the Euler number reads $\eta = 2( h^{1,1} - h^{1,2})$. For four-folds, the non-trivial Hodges are $h^{1,1}$, $h^{1,2}$, $h^{1,3}$, $h^{2,2}$, with the Euler number given by $\eta = 4 + 2 h^{1,1}-4 h^{1,2}+2 h^{1,3}+h^{2,2}$; the associated plot is shown in the top-right. A detailed analysis on the properties of the cohomological data of CICY four-folds can be found in \cite{Gray_2014}. Finally, we represent the manifolds from complex dimension three, four and five on a single plot, by employing logarithmic axes.
This is shown in the bottom plot, where an additional line is drawn to indicate how the asymptotic behaviour of three-folds and five-folds exactly match. They both cluster along the line\footnote{Just for reference, we observe that the four-folds data points are approximately bounded by the line $y = 5/6 \, x +4$.}
\begin{align}
   y  = -\frac{1}{2} x + 4 \, .
\end{align}
This analysis shows one of the interesting investigations that naturally follow from our results. The knowledge of the holomorphic invariants for five-folds allows to study Calabi-Yau properties across the first two non-trivial odd complex dimensions. This is subject of ongoing work, aimed at unveiling such a connection.

\section{Neural Network analysis}
\label{sec:NN}
In this section, we describe the neural network architectures that we employed for the investigations on the dataset. We were inspired by some recent applications of ML techniques to the lower-dimensional CICY datasets \cite{ Bull:2018uow, Bull:2019cij,Erbin_2021, Erbin_2021_m, Erbin_2022, He:2020lbz}.

Before doing that, we list some choices concerning the analysis of the data that we present in this section. Regarding the splitting into training, validation, test, we opted for a split into $70\%$, $20 \%$, $10 \%$ (in order). All matrices have been padded to the maximum dimension of $4 \times 4$. The batch size was set to $128$. All the investigations were run on a Macbook Air M1 with $16$GB RAM and with $8$ cores. We employed PyTorch, which uses the Metal Performance Shaders (MPS) backend for GPU training acceleration. 

\subsection{Classifier}
A classifier neural network is most efficient when the number of classes is much smaller than the size of the dataset, and the samples are distributed somehow uniformly among them. For these reasons, this kind of architecture is particularly suitable for learning $h^{1,1}$. The most natural choice, following \cite{He:2020lbz}, can be summarised as:
\begin{align}
    N_{4*4}(512, \sigma , \delta_{0.4}, 256, \sigma , \delta_{0.3}, 256, \sigma , N_h, S),
    \label{eq:Classifier_h11}
\end{align}
where $N_h$ is the number of possible $h^{1,1}$ values, i.e. the number of classes. In our case we have that $N_h=16$, which gives a total of $209936$ parameters. For whole CICY four-folds dataset, where $N_h = 24$, the authors in \cite{He:2020lbz} report a $96\%$ accuracy, with $20\%$ ($15+5$) of dataset used for the training session. Motivated by their results, we employ this architecture for our investigations on $h^{1,1}$. 

Regarding the other Hodge numbers, we modify the above architecture slightly, to the following neural network:
\begin{align}
    N_{4*4}(512, \sigma , \delta_{0.4}, 512, \sigma , \delta_{0.3}, 512 , \sigma , \delta_{0.3}, 256, \sigma , N_H, S),
    \label{eq:Classifier_2}
\end{align}
where $N_H$ is the number of classes for the higher Hodge numbers (which can be found in table \ref{tab:my_label}).

\subsection{Linear Regressor}
Moving to the realm of regressors, we begin with those of the simplest type, i.e. linear regressors. Since such architectures do not rely on a fixed number of classes, they can be more easily adapted to different problems. Again, by following the work presented in \cite{He:2020lbz}, we take the network
\begin{align}
    N_{4*4}(512, s, 256, s, 128, s, 32, s, 8, s, 1)
    \label{eq:Regressor_h11}
\end{align}
as the starting point for learning $h^{1,1}$. It is composed of $177329$ parameters. For the four-fold case, the accuracy reached by this network is roughly $93\%$. We use exactly this network for investigating $h^{1,1}$ in our case. \\
For higher Hodge numbers, we
employ the slightly larger neural network
\begin{align}
    N_{4*4}(1024, s, 1024, s, 512, s, 64, s, 16, s, 1),
\label{eq:Regressor_higher_h}
\end{align}
with $1625697$ parameters.

\subsection{Convolutional Regressor}
The last type of architecture that we employed for our investigations is the convolutional regressor. This choice is motivated by the results in \cite{Erbin_2021_m}. Specifically, the neural network described therein can be written as
\begin{align}
    N_{4*4}(C_{5*5}180, r, BN_{0.99}, C_{5*5}100, r, BN_{0.99}, C_{5*5}40, r, BN_{0.99}, C_{5*5}20, r, BN_{0.99}, \delta_{0.4}, f, 1, r),
    \label{eq:Convolutional}
\end{align}
involving $575841$ trainable parameters.
The results of this network in the prediction of $h^{1,1}$ for three-folds by the authors above has an accuracy of $94\%$. We test this on all the Hodge numbers

\section{Results}
\label{sec:Results}
In this section, we present the results of our ML investigations on the cohomological and topological properties contained in the dataset with the architectures just described.
We mainly focus on two measures: the standard $R^2$ score and the accuracy, defined as:
\begin{align}
    \tx{accuracy} = \tx{correct predictions / total predictions} \, .
\end{align}

\subsection{\texorpdfstring{$h^{1,1}$}{Lg}}
It is generally the case that $h^{1,1}$ is learnt with higher precision and accuracy, compared to the other Hodge numbers (see \cite{Erbin_2021_m, Erbin_2022, He:2020lbz, Berman_2022, Ed}). Our investigation is no exception; we find that, by simply applying the architectures that were devised for CICY four-folds, very good results are obtained. They are shown in table \ref{tab:h11} below.
\begin{table}[H]
    \centering
    \begin{tabular}{|c|c|c|}
    \cline {2 - 3}
      \multicolumn{1}{c|}{} & $\quad R^2 \quad$ & Accuracy \\
       \hline 
     Classifier (\ref{eq:Classifier_h11}) & n/a &  $88 \%$ \\
       \hline
       Linear Regressor (\ref{eq:Regressor_h11}) &  $85 \%$ &  $93 \%$ \\
       \hline Convolutional Regressor (\ref{eq:Convolutional}) &  $91 \%$ &  $96 \%$ \\ \hline
    \end{tabular}
    \caption{ML performances on the Hodge number $h^{1,1}$.}
    \label{tab:h11}
\end{table}
This is a very promising result, showing that for CICY five-folds, in line with the results for three-folds and four-folds, machine learning is very effective in predicting the lowest non-trivial Hodge number. One of the reasons for this performance is the limited range of $h^{1,1}$, shown in figure \ref{fig:Histograms}.

\subsection{\texorpdfstring{$h^{1,2}$}{Lg}}
As it can be seen in figure \ref{fig:Histograms}, the distribution of this Hodge number is heavily skewed towards zero, because the vast majority of the $h^{1,2}$'s vanish. Probably due to these features, we find that the neural network struggles in the training process, which is evident from the negative $R^2$ measure. This is shown in Table \ref{tab:h12}.
\begin{table}[H]
    \centering
    \begin{tabular}{|c|c|c|}
    \cline {2 - 3}
      \multicolumn{1}{c|}{} & $\quad R^2 \quad$ & Accuracy \\
       \hline 
     Classifier (\ref{eq:Classifier_2}) & n/a &  $98 \%$ \\
       \hline
       Linear Regressor (\ref{eq:Regressor_higher_h}) &  -ve &  $83 \%$ \\
       \hline Convolutional Regressor (\ref{eq:Convolutional}) &  -ve &  $97 \%$ \\ \hline
    \end{tabular}
    \caption{ML performances on the Hodge number $h^{1,2}$.}
    \label{tab:h12}
\end{table}
In fact, $h^{1,2}$ is zero $97.7 \%$ of the times, which makes the distribution of the samples not suitable for ML purposes, since the network does not have enough non-zero values for training. This percentage is also showing that, as we expect, the classifier is doing no better than guessing $h^{1,2}=0$ by default.

\subsection{\texorpdfstring{$h^{1,3}$}{Lg}}
The distribution of $h^{1,3}$ is similar to the one of $h^{1,2}$, with slightly less vanishing numbers. We observe a slight improvement in the $R^2$ score, but the inadequateness of the samples, discussed in the previous case, 
still applies. This is illustrated in Table \ref{tab:h13}.
\begin{table}[H]
    \centering
    \begin{tabular}{|c|c|c|}
    \cline {2 - 3}
      \multicolumn{1}{c|}{} & $\quad R^2 \quad$ & Accuracy \\
       \hline 
     Classifier (\ref{eq:Classifier_2}) & n/a &  $96 \%$ \\
       \hline
       Linear Regressor (\ref{eq:Regressor_higher_h}) &  $61 \%$ &  $93 \%$ \\
       \hline Convolutional Regressor (\ref{eq:Convolutional}) &  $79 \%$ &  $95 \%$ \\ \hline
    \end{tabular}
    \caption{ML performances on the Hodge number $h^{1,3}$.}
    \label{tab:h13}
\end{table}
We find that $95.7 \%$ of $h^{1,3}$'s vanish, showing again that the classifier is probably always predicting the class $h^{1,3}=0$.

\subsection{\texorpdfstring{$h^{1,4}$}{Lg}}
In this case, the distribution is very different from the previous two discussed. $h^{1,4}$ can take a much larger range of values, which are scattered throughout 3 order of magnitudes. We first present the results, in table \ref{tab:h14}, then comment on their interpretation.
\begin{table}[H]
    \centering
    \begin{tabular}{|c|c|c|c|c|}
    \cline {2 - 4}
      \multicolumn{1}{c|}{} & $\quad R^2 \quad$ & Accuracy & Accuracy w/ $10\%$ Tolerance \\
       \hline 
     Classifier (\ref{eq:Classifier_2}) & n/a &  $3 \%$ & n/a\\
       \hline
       Linear Regressor (\ref{eq:Regressor_higher_h}) &  $98 \%$ &  $3 \%$ & $86 \%$ \\
       \hline Convolutional Regressor (\ref{eq:Convolutional}) &  $98 \%$ &  $2 \%$ & $86\%$\\ \hline
    \end{tabular}
    \caption{ML performances on the Hodge number $h^{1,4}$.}
    \label{tab:h14}
\end{table}
These results hint at a very simple behaviour: the neural network is learning the pattern behind the Hodge number (hence the very high $R^2$ score), but is not able to predict the exact integer. It is a reasonable outcome, since the range is very large, as we mentioned, and the size of the training set is probably not optimal. However, despite the failure to give exact predictions, the network is providing good approximations. To give a quantitative reference for this, we tested an additional measure, which is reported in the last column of the table. The \textit{accuracy with 10}$\%$ \textit{tolerance} counts how many predictions differ from the actual value by less than $10 \%$ of it, and divides their number by the total number of predictions. Hence, the results show that the regressor neural networks are indeed able to approximate the Hodge numbers well, despite they do not manage to predict their exact values.

\subsection{\texorpdfstring{$h^{2,2}$}{Lg}}
The distribution of $h^{2,2}$ somehow interpolates between the one of $h^{1,2/3}$ and the one of $h^{1,4}$; most of the values are close to zero (even though only $5$ of them are actually zero) and the upper bound is $510$, giving it quite a wide range. We find that the machine learning performances are quite poor with respect to both measures, as shown in table \ref{tab:h22} below.
\begin{table}[H]
    \centering
    \begin{tabular}{|c|c|c|}
    \cline {2 - 3}
      \multicolumn{1}{c|}{} & $\quad R^2 \quad$ & Accuracy \\
       \hline 
     Classifier (\ref{eq:Classifier_2}) & n/a &  $22 \%$ \\
       \hline
       Linear Regressor (\ref{eq:Regressor_higher_h}) &  $26 \%$ &  $25 \%$ \\
       \hline Convolutional Regressor (\ref{eq:Convolutional}) &  $16 \%$ &  $27 \%$ \\ \hline
    \end{tabular}
    \caption{ML performances on the Hodge number $h^{2,2}$.}
    \label{tab:h22}
\end{table}

\subsection{\texorpdfstring{$h^{2,3}$}{Lg}}
Our findings show that $h^{2,3}$ has the largest number of possible values and it spans the widest range among the Hodge numbers. Its distribution is very similar to the one of $h^{1,4}$, scaled by a factor of $10$. Accordingly, we observe similar results, with very high $R^2$ measure but very low accuracy, shown in table \ref{tab:h23}. 
\begin{table}[H]
    \centering
    \begin{tabular}{|c|c|c|c|}
    \cline {2 - 4}
      \multicolumn{1}{c|}{} & $\quad R^2 \quad$ & Accuracy & Accuracy w/ $10\%$ Tolerance \\
       \hline 
     Classifier (\ref{eq:Classifier_2}) & n/a &  $2.7 \%$ & n/a \\
       \hline
       Linear Regressor (\ref{eq:Regressor_higher_h}) &  $97 \%$ &  $0.2 \%$ & $77\%$\\
       \hline Convolutional Regressor (\ref{eq:Convolutional}) &  $98 \%$ &  $0.1 \%$ & $85\%$\\ \hline
    \end{tabular}
    \caption{ML performances on the Hodge number $h^{2,3}$.}
    \label{tab:h23}
\end{table}
Again, by employing the accuracy with $10 \%$ tolerance measure, introduced for the $h^{1,4}$ investigation, we show that the neural network is able to approximate the Hodge numbers well.

\subsection{\texorpdfstring{$\eta$}{Lg}}
We end this section by presenting the investigation on the Euler number. Its range is again quite extended, which makes classification efforts not effective, as it can be seen from table \ref{tab:eta} below.
\begin{table}[H]
    \centering
    \begin{tabular}{|c|c|c|c|}
    \cline {2 - 4}
      \multicolumn{1}{c|}{} & $\quad R^2 \quad$ & Accuracy & Accuracy w/ $10\%$ Tolerance \\
       \hline 
     Classifier (\ref{eq:Classifier_2}) & n/a &  $3 \%$ & n/a \\
       \hline
       Linear Regressor (\ref{eq:Regressor_higher_h}) &  $50 \%$ &  $0 \%$ & $0\%$\\
       \hline Convolutional Regressor (\ref{eq:Convolutional}) &  $98 \%$ &  $0.04 \%$ & $83\%$\\ \hline
    \end{tabular}
    \caption{ML performances on the Hodge number $\eta$.}
    \label{tab:eta}
\end{table}
In this case, the linear regressor also gives very poor results, with a low $R^2$ score and zero accuracies. The convolutional regressor, on the other hand, is still performing as a good approximator, with very high $R^2$ and a promising $83 \%$ accuracy with $10\%$ tolerance.

\section{Conclusions}
\label{sec:Conclusions}
We presented a partial classification of complete intersection Calabi-Yau (CICY) five-folds and their Hodge diamonds. Due to their importance in compactification of F-theory and M-theory, these spaces have appeared multiple times in the physics literature in the last twenty years. Lower-dimensional CICY's, with analogous applications in the dimensional reduction of string theory, have already been classified in previous works, and their Hodge numbers have been computed. The key mathematical fact that was employed for the computations in those cases is the existence of the short exact \textit{adjunction sequence}. However, in complex dimension five, this is not sufficient to determine the complete Hodge diamond of CICY's. We showed that, by also considering the symmetrised version of the above sequence, it is possible to determine the full Hodge diamond. On the other hand, the generation of configuration matrices does not differ from the lower-dimensional analogues. Both procedures (the generation of configuration matrices and the computation of Hodge numbers) were implemented on a computer, and applied to a subset of all the possible spaces. This subset was chosen according to the following results and considerations. We derived the maximum size for the configuration matrix, which is $25 \times 30$, and estimated the expected size of the dataset to be in the order of $10^8$. Since this number, with our current algorithm, makes the construction of the whole dataset unfeasible, it was natural to restrict ourselves to a subset of it. Motivated by the recent machine learning investigations presented in \cite{He:2020lbz}, we chose to focus on all configuration matrices with size up to $4 \times 4$, a subset that yielded promising results in the above work based on four-folds. We obtained $27068$ matrices, inequivalent under permutations of rows and columns, for all of which we calculated the corresponding Euler number. Out of these, $3909$ were excluded from the rest of our investigations, because they represent product spaces. Of the remaining $23159$ non-product matrices, we computed the full Hodge diamond for $12433$ of them ($53.7\%$), finding  $2375$ different sets of Hodge numbers.\footnote{The dataset can be downloaded \href{https://www.dropbox.com/scl/fo/z7ii5idt6qxu36e0b8azq/h?rlkey=0qfhx3tykytduobpld510gsfy&dl=0}{\textcolor{blue}{here}}.} These figures are large enough for machine learning techniques to be effective. Hence, we attempted to learn all the non-trivial Hodge numbers, namely $h^{1,1}$, $h^{1,2}$, $h^{1,3}$, $h^{1,4}$, $h^{2,2}$,$h^{2,3}$, and the Euler number $\eta$, with three types of architectures: classifier, linear regressor and convolutional regressor. In agreement with the machine learning investigations on the lower-dimensional CICY datasets, we find that $h^{1,1}$ can be learnt to very high levels of accuracy, with the best result being achieved by the convolutional regressor. For this architecture, we find an $R^2$ score of $91 \%$ and an accuracy, defined as the ratio of the exact predictions to the total predictions, of $ 96\%$. This result corroborates what was found by analogous investigations for three-folds and four-folds, and the success of ML is partially due to the fact that the range of $h^{1,1}$ is relatively small. The opposite is true for $h^{1,4}$, $h^{2,3}$ and $\eta$, with each of their distributions spanning many order of magnitudes. This makes any classification efforts almost hopeless, but it is still sensible to apply regressor neural networks. For those, we find very high $R^2$ scores for the invariants above, but, again due to the size of the range, only a very small fraction of the predicted values exactly match the Hodge numbers. However, we find that around $85 \%$ of the predictions differ by less than $10 \%$ from the exact result for the convolutional regressor, showing it to be an efficient approximator. Due to their distributions, the remaining Hodge numbers did not yield promising results. A more complete set of samples would be beneficial for a proper training of the neural network.
This brings us to the final considerations about this work, and its possible extensions. 

The construction of the full CICY five-folds dataset, which is the most natural development of our work, is itself a very challenging task. As we mentioned, because of its estimated size in the order of $10^8$, building it completely is likely be very heavy computationally. Machine
learning could help make this process more manageable, and could be useful to quickly extract information on the full dataset, such as bounds for the Hodge numbers, by just constructing part of it and predicting the rest of it. Progresses in this direction, i.e. extrapolating predictions from low to high Hodge numbers, have been made for CICY three-folds and four-folds in \cite{Erbin_2023}.
This new collection of data would also allow for an investigation of the properties of CICY's cohomological numbers across different complex dimensions, both with and without neural networks. It is something that has never been done before, and it has the potential unveil unknown properties and reveal a new predictive power. \\
Moreover, we believe that substantial improvements on the ML performance could be reached also with the current subset of data. Applying the inception network developed for CICY four-folds in \cite{Erbin_2022} would be a promising starting point in that direction. \\
Finally, another interesting extension consists of a detailed analysis of the topological properties in our dataset. This would include computing the intersection numbers, and other investigations along the lines of \cite{Gray_2014}. Such a study would also have a particular relevance from the phenomenological perspective, in that it would identify the elliptically fibered spaces.

\section{Acknowledgements}
We are grateful to Professor Giorgio Ottaviani for useful discussions on the spectral sequence techniques involved. TSG would like to thank Professor David Berman for his support during the early stages of this project. DA was supported by project PRIN2022 "Real and Complex Manifolds: Geometry and
Holomorphic Dynamics" (code 2022AP8HZ9) and by GNSAGA of
INdAM. TSG was supported by the Science and Technology Facilities Council (STFC) Consolidated Grants ST/T000686/1 "Amplitudes, Strings \& Duality" and ST/X00063X/1 "Amplitudes, Strings \& Duality". New data were generated and analysed during this study.

\appendix

\section{Appendix}

\subsection{Finiteness of the CICY Five-folds}
In this section, we present the proof of the bounds on the configuration matrices stated in section \ref{sec:Size}. \\
The first step is to isolate the $\mathbb{CP}^1$ factors in the ambient space, and we do so by adopting the following notation:
\begin{align}
    \mathcal{A} = (\mathbb{CP}^1)^f \times \mathbb{CP}^{n_1} \times \dots \times \mathbb{CP}^{n_F} \, .
\end{align}
Since we have $f$ projective spaces of dimension $1$ and $F$ projective spaces of dimensions $n_j>1$ ($j=f+1,...,f+F)$, it is clear that $\mathrm{dim}(\mathcal{A})=f + \sum_{j=f+1}^{f+F} n_j$. Then, according to \eqref{eq:K_and_n}, we need $K$ constraints such that
\begin{align}
    f + \sum_{j=f+1}^{f+F} n_j - K = 5
    \label{eq:f_K_5}
\end{align}
to construct a \textit{five}-fold. The Calabi-Yau condition \eqref{eq:CY_Condition} translates into:
\begin{align}
    &\sum_{\alpha=1}^K q_{\alpha}^j = 2 \quad \quad \mathrm{for }\,\, j = 1,...,f  \, ,\nonumber \\
    &\sum_{\alpha=1}^K q_{\alpha}^{j} = n_j + 1 \quad \quad \mathrm{for }\,\, j = f+1,...,f + F \, .
\end{align}
Without loss of generality, we can assume that all constraints obey:\footnote{Suppose that there is a constraint which is trivial everywhere but on a single projective space, with degree one. Then, if we remove the corresponding column from the configuration matrix and modify accordingly one of the other constraints, we obtain a configuration matrix describing the same space.}
\begin{align}
    \sum_{j=1}^{f+F} q_{\alpha}^j \geq 2 \, .
    \label{eq:Cons_bigger_than_two}
\end{align}
From the equations above, we obtain
\begin{align}
2 K \leq \sum_{\alpha=1}^K \sum_{j=1}^{f+F} q_{\alpha}^j=2 f+\sum_{j=f + 1}^{f+F}\left(n_j+1\right) = f + K + 5 + F\, ,
\end{align}
which gives:
\begin{align}
    F + f + 5 \geq K \, .
\end{align}
Let us now find an upper bound for the quantity $\sum_{j = f+1}^{f +F} (n_j -1)$. By using \eqref{eq:f_K_5} again, and the above inequality, we get:
\begin{align}
    \sum_{j = f+1}^{f +F} (n_j -1) = \sum_{j = f+1}^{f +F} n_j - F = K +5 - f - F \leq 10 \, .
\end{align}
This allows us to infer the following inequality:
\begin{align}
    10 \geq \sum_{j = f+1}^{f +F} (n_j -1) \geq F \, ,
\end{align}
and also
\begin{align}
    \sum_{j = f+1}^{f +F} n_j \leq 10 + F \leq 20 \, .
    \label{eq:N_j_smaller_than_twenty}
\end{align}
Hence, we have just derived an upper bound for the number of projective spaces with dimensions greater than one. To determine the same constraint for the number of $\mathbb{CP}^1$'s, we need to make a few considerations on the equivalence between certain configuration matrices. Let us first observe that 
\begin{align}
\left(\begin{array}{c|cc}
1 & 1 & a \\
1 & 1 & b \\
X & 0 & M
\end{array}\right) \simeq\left(\begin{array}{c|c}
1 & a+b \\
X & M
\end{array}\right),
\label{eq:Equiv_conf_mat_q}
\end{align}
for generic $a,b,X,M$. Hence, we can assume that any bilinear constraint that involves a $\mathbb{CP}^1$ factor must involve also a $\mathbb{CP}^{n_j}$ factor with $j \in \{ f+1, ..., f+F \}$ (if it were to involve two $\mathbb{CP}^1$ factors, we could apply the equivalence above). Let us assume that there are $t$ constraints of this type, which we denote as
\begin{align}
Q_{\alpha}^j = q_{\alpha}^{j} \, \, \mathrm{s.t.} \,\,
\sum_{j=1}^f q_{\alpha}^j=1, \quad \sum_{j=f+1}^F q_{\alpha}^j=1 \, .
\end{align}
All these constraints involve one of the $\mathbb{CP}^{n_j}$ factors with $j \in \{ f+1, ..., f+F \}$, hence we have that
\begin{align}
t \leq \sum_{\alpha=1}^K \sum_{j=f+1}^{f+F} q_{\alpha}^{j} = \sum_{j=f+1}^{f+F}\left(n_j+1\right)=2 F + \sum_{j = f+1}^{f +F} (n_j -1) \, .
\end{align}
Let us make another remark on the general form of the configuration matrices of the form \eqref{eq:Equiv_conf_mat_q} (left hand side). Suppose that the constraint on the left is again of degree two. This time, assume that it is of degree two over one $\mathbb{CP}^1$, and therefore trivial everywhere else. This implies that the configuration matrix is block diagonal, having the form of a product space. Hence, when considering a constraint with degree bigger than one over the one-dimensional projective spaces, labelled by some $\alpha$, we impose
\begin{align}
    \sum_{j=1}^f q_{\alpha}^j > 2 \, .
    \label{eq:Bigger_than_three}
\end{align}
This ensures that more than one $\mathbb{CP}^1$ factor is involved.
We denote the sum of the degrees over the $\mathbb{CP}^1$ factors coming from all constraints of this type with $s$. Then, by definition, we have that $t+s = 2f$. \\
Now, let us focus on the following quantity:
\begin{align}
    \begin{aligned}
& \sum_{\alpha=1}^K\left(\sum_{j=1}^{f+F} q_{\alpha}^j-2\right)=\sum_{\alpha=1}^K \sum_{j=1}^{f+F} q_a^j-\sum_{\alpha=1}^K 2 = \big[2 f+2 F+ \sum_{j = f+1}^{f +F} (n_j -1) \big]-2 K \\
& = \big[2 f+2 F+ \sum_{j = f+1}^{f +F} (n_j -1) \big] - 2 \big[ f + F + \sum_{j = f+1}^{f +F} (n_j -1) - 5 \big]= 10 -\sum_{j = f+1}^{f +F} (n_j -1) \, .
\end{aligned}
\end{align}
In the light of \eqref{eq:Cons_bigger_than_two}, if all the constraints in the configuration have degree smaller or equal to three, then this quantity counts the number of constraints with degree three. Let us outline what happens with constraints with higher degree. Any constraint of degree $g>3$, together with $g-3$ constraints of degree two, can be exchanged for $g-2$ constraints of degree three. In order to find which of the two cases (before or after the exchange) maximises $s$, we look at both. In the first case, the maximum value of $s$ is $g$ (recall the condition \eqref{eq:Bigger_than_three}). In the second case, $s$ can be up to $3(g-2)$, which is greater that $g$ since $g>3$. Hence, the maximum value of $s$ is attained when all the constraints have degree 3, which yields
\begin{align}
    s \leq 3 \big[ 10 - \sum_{j = f+1}^{f +F} (n_j -1) \big] \, .
\end{align}
This implies that
\begin{align}
    &f = \frac{t+s}{2} \leq \frac{2 F + \sum_{j = f+1}^{f +F} (n_j -1) + 3 \big[ 10 - \sum_{j = f+1}^{f +F} (n_j -1) \big] }{2} = \nonumber \\
    & = 15 + F - \sum_{j = f+1}^{f +F} (n_j -1) \leq 15 \, .
\end{align}
Hence, this gives $f+F \leq 25$. Since we had that $ \sum_{j = f+1}^{f +F} n_j \leq 20$ (see \eqref{eq:N_j_smaller_than_twenty}), then an upper bound for the dimensions of the ambient space is given by 
\begin{align}
    \mathrm{dim}(\mathcal{A}) = f + \sum_{j = f+1}^{f +F} n_j \leq 35 \, .
\end{align}
According to \eqref{eq:f_K_5}, the maximum number of constraints is therefore $30$.
Summarising, in complex dimension five we have that:
\begin{align}
\mathrm{the \,\, maximum \,\, size \,\, of \,\, a \,\,  configuration \,\, matrix \,\, describing  \,\,a \,\, CICY\,\, is} \, \, 25 \times 30 \, .
\end{align}

\subsection{Generation Algorithm Exemplified}
In this section, we present an example that illustrates the steps involved in the algorithm for the generation of inequivalent matrices. \\
Let us choose $3\times3$ as the maximum dimension of the configuration matrix, by which we mean that $q_{\alpha}^r$ should be at most $3 \times 3$. Then, the algorithm finds the possible ambient spaces, $\mathbf{n}$, which satisfy this condition, which are:
\begin{gather}
m = 1:
\left(\begin{array}{c}
  6
\end{array}\right| ,
\left(\begin{array}{c}
  7
\end{array}\right| ,
\left(\begin{array}{c}
  8
\end{array}\right| .  \nonumber \\
m = 2:  
\left(\begin{array}{c}
   1 \\
   5 
\end{array}\right| ,
\left(\begin{array}{c}
   2 \\
   4 
\end{array}\right| ,
\left(\begin{array}{c}
   3 \\
   3 
\end{array}\right| ,
\left(\begin{array}{c}
   1 \\
   6 
\end{array}\right| ,
\left(\begin{array}{c}
   2 \\
   5 
\end{array}\right| ,
\left(\begin{array}{c}
   3 \\
   4 
\end{array}\right| ,
\left(\begin{array}{c}
   1 \\
   7
\end{array}\right| ,
\left(\begin{array}{c}
   2 \\
   6 
\end{array}\right| ,
\left(\begin{array}{c}
   3 \\
   5 
\end{array}\right| ,
\left(\begin{array}{c}
   4 \\
   4 
\end{array}\right| . \nonumber \\
m = 3:  
\left(\begin{array}{c}
   1 \\
   1  \\
   4
\end{array}\right| ,
\left(\begin{array}{c}
   1 \\
   2 \\
   3 
\end{array}\right| ,
\left(\begin{array}{c}
   2 \\
   2 \\
   2
\end{array}\right| ,
\left(\begin{array}{c}
   1 \\
   1  \\
   5
\end{array}\right| ,
\left(\begin{array}{c}
   1 \\
   2 \\
   4 
\end{array}\right| ,
\left(\begin{array}{c}
   1 \\
   3 \\
   3
\end{array}\right| ,
\left(\begin{array}{c}
   2 \\
   2  \\
   3
\end{array}\right| , 
\left(\begin{array}{c}
   1 \\
   1 \\
   6 
\end{array}\right| , \nonumber \\
\left(\begin{array}{c}
   1 \\
   2 \\
   5
\end{array}\right| , 
\left(\begin{array}{c}
   1 \\
   3  \\
   4
\end{array}\right| ,
\left(\begin{array}{c}
   2 \\
   2 \\
   4 
\end{array}\right| ,
\left(\begin{array}{c}
   2 \\
   3 \\
   3
\end{array}\right| . 
\label{eq:List_of_ns}
\end{gather}
Let us pick one of these to illustrate the successive step. To fit this example within a page or so, let us choose $\mathbf{n} = \left(\begin{array}{c}
  2 \\
  2 \\
  3
\end{array}\right| $. \\
The algorithm now finds the sets of rows satisfying the Calabi-Yau condition \eqref{eq:CY_Condition}, but it does so while also taking into account that some projective spaces may be identical. Specifically, the first and second factors are the same, and this is taken into account to avoid redundant elements in the set. Hence, we generate all valid inequivalent polynomials living in the first two ambient spaces, and all valid inequivalent polynomials living in the last ambient space; once arranged in lexicographic order, the list reads:
\begin{align}
\Big\{\, \, \big[ \, \left|\begin{array}{cc}
    0 & 3 \\
    0 & 3
\end{array}\right) \, , \,\,  \left|\begin{array}{cc}
    0 & 3 \\
    1 & 2
\end{array}\right) \, , \,\,  \left|\begin{array}{cc}
    1 & 2 \\
    1 & 2
\end{array}\right) \big]\, , \, \, \big[ \left|\begin{array}{cc}
    0 & 4
\end{array}\right) \, , \,\, \left|\begin{array}{cc}
    1 & 3
\end{array}\right) \, , \,\, \left|\begin{array}{cc}
    2 & 2
\end{array}\right)  \big] \,\, \Big\}.
\end{align}
The next step consists of taking the Cartesian product of the above sets, in order to find all the possible configuration matrices out of those. We find:
\begin{gather}
\left|\begin{array}{cc}
    0 & 3 \\
    0 & 3 \\
    0 & 4
\end{array}\right) , \,
\left|\begin{array}{cc}
    0 & 3 \\
    0 & 3 \\
    1 & 3
\end{array}\right) , \,
\left|\begin{array}{cc}
    0 & 3 \\
    0 & 3 \\
    2 & 2
\end{array}\right) , \,
\left|\begin{array}{cc}
    0 & 3 \\
    1 & 2 \\
    0 & 4
\end{array}\right) , \,
\left|\begin{array}{cc}
    0 & 3 \\
    1 & 2 \\
    1 & 3
\end{array}\right) , \,
\left|\begin{array}{cc}
    0 & 3 \\
    1 & 2 \\
    2 & 2
\end{array}\right) , \nonumber \\
\left|\begin{array}{cc}
    1 & 2 \\
    1 & 2 \\
    0 & 4
\end{array}\right) , \,
\left|\begin{array}{cc}
    1 & 2 \\
    1 & 2 \\
    1 & 3
\end{array}\right) , \,
\left|\begin{array}{cc}
    1 & 2 \\
    1 & 2 \\
    2 & 2
\end{array}\right) \, ,
\label{eq:List_of_matrices}
\end{gather}
where, again, we ordered the matrices lexicographically. Note that, since we keep the lexicographic order in every step, the matrices generated at this stage cannot be related by permutation of rows and columns, by construction.\footnote{This is evident from the fact there are no pairs of matrices sharing the same elements.}
Now, we have to pick one of them in order to proceed to the next step, and (for illustrative purposes) we choose 
\begin{align}
    q_{\alpha}^r = \left|\begin{array}{cc}
    1 & 2 \\
    1 & 2 \\
    0 & 4
\end{array}\right) \, .
\end{align}
For this matrix, we perform all permutations within the same row, again avoiding repetitions when two rows are identical, and we obtain:
\begin{gather}
\left|\begin{array}{cc}
    1 & 2 \\
    1 & 2 \\
    4 & 0
\end{array}\right) \, , \,\, \left|\begin{array}{cc}
    1 & 2 \\
    1 & 2 \\
    0 & 4
\end{array}\right) \, , \,\, 
\left|\begin{array}{cc}
    1 & 2 \\
    2 & 1 \\
    4 & 0
\end{array}\right) \, , \,\,
\left|\begin{array}{cc}
    1 & 2 \\
    2 & 1 \\
    0 & 4
\end{array}\right) \, , \,\,
\left|\begin{array}{cc}
    2 & 1 \\
    2 & 1 \\
    4 & 0
\end{array}\right) \, , \,\, 
\left|\begin{array}{cc}
    2 & 1 \\
    2 & 1 \\
    0 & 4
\end{array}\right) .
\end{gather}
The above set can contain matrices that are related by permutations of rows and columns, and this is checked using the algorithm of \cite{Gray_2013}, with brute force comparison when the eigenvalues are degenerate. This yields only three inequivalent matrices:
\begin{align}
    \left|\begin{array}{cc}
    1 & 2 \\
    1 & 2 \\
    4 & 0
\end{array}\right) \, , \,\, \left|\begin{array}{cc}
    1 & 2 \\
    1 & 2 \\
    0 & 4
\end{array}\right) \, , \,\, 
\left|\begin{array}{cc}
    1 & 2 \\
    2 & 1 \\
    4 & 0
\end{array}\right) \, .
\end{align}
Now, the same steps must be performed for the other matrices listed in \eqref{eq:List_of_matrices}, before moving to another $\mathbf{n}$ from \eqref{eq:List_of_ns}.

\subsection{Hodge Number Computation Exemplified}
We provide the explicit steps of the computation outlined in sections \ref{sec:Adjunction}, \ref{sec:Sym_adjunction}, \ref{sec:Algorithm_Hodge} for a specfic example: the CICY five-fold described by
\begin{align}
    \left(\begin{array}{c|ccc}
2 & 1 & 1 & 1 \\
6 & 0 & 0 & 7
\end{array}\right) .
\label{eq:Ex_configur_mat}
\end{align}
As we mentioned in the relevant section, general formulae for the first page in the spectral sequences associated to the tangent bundle and normal bundle have been presented in \cite{article_Green} (as well as in \cite{Green:1987cr}). We refer the reader to the above article, together with the more pedagogical treatment in \cite{Hubsch:1992nu}, for more details. We just report the main definitions and results here. Let us first clarify the notation. Given a holomorphic vector bundle $\mathcal{V}$ over $\mathcal{A}$, we denote the $i^{th}$ page with $E_i^{j,k}(\mathcal{V})$, and the $i^{th}$ differential is of the form
\begin{align}
    d_i: E_i^{j, k}(\mathcal{V}) \rightarrow E_i^{j-i+1, k-i}(\mathcal{V}) \, .
\end{align}
The first page is defined as:
\begin{align}
    E_1^{j,k}(\mathcal{V}) = H^j ( \mathcal{A}, \wedge^k \mathcal{E}^* \otimes \mathcal{V} ) \, ,
\end{align}
with the subsequent ones being $E_{r+1}(\mathcal{V}) = H(E_r(\mathcal{V}), d_r)$. For our problem, we are interested in the special cases of $\mathcal{V}$ being $\mathcal{E}$ and $\mathcal{T}_{\mathcal{A}}$. As we mentioned, the first pages of both sequences have already been worked out in \cite{article_Green} (or, alternatively, in \cite{Green:1987cr}), and they read:
\begin{align}
E_1^{j, k}\left(\mathcal{T}_{\mathcal{A}}\right)= & \bigoplus_{|A|=k} \bigoplus_{r=1}^m \bigoplus_{\Sigma \gamma_t=j}\left[ H^{\gamma_r}\left(\mathbb{C P}^{n_r}, \mathcal{T}(\mathbb{CP}^{n_r}) \otimes\left(h_r\right)^{-\sum_{\alpha \in A} q_{\alpha}^r}\right)\right. \nonumber \\
& \bigotimes_{\substack{s=1 \\
s \neq r}}^m H^{\gamma_s}\left(\mathbb{C P}^{n_s},\left(h_s\right)^{-\sum_{\alpha \in A} q_{\alpha}^s}\right)\Big] ,
\end{align}
for the tangent bundle of the ambient space, and
\begin{align}
E_1^{j,k}(\mathcal{E}) = 
    \bigoplus_{\beta=1}^{K} \bigoplus_{\Sigma \gamma_t=j} \bigoplus_{\substack{|A|=k \\ b \notin A}} \left[ \bigotimes_{s=1}^{m} H^{\gamma_s}\left(\mathbb{C P}^{n_s}, \left(h_s \right)^{ q_{\beta}^s -\sum_{\alpha \in A} q_{\alpha}^s}\right)\right],
\end{align}
for the normal bundle. In both equations $A$ denotes a subset of indices $\alpha = 1, ..., K$, and $|A|$ its cardinality. 
To calculate the dimensions of the cohomologies involved, we use Bott's formula:
\begin{align}
&\operatorname{rank} H^q\left(\mathbb{CP}^n, \Lambda^p \mathcal{T}\left(\mathbb{CP}^n\right) \otimes h^k\right) = \vspace{0.3cm} \nonumber  \\
&=\left\{\begin{array}{lll}
\left(\begin{array}{c}
k+n+1+p \\
p
\end{array}\right)\left(\begin{array}{c}
k+n \\
n-p
\end{array}\right) & \tx{if} \quad q=0 \text { and } k \geqslant-p &  \vspace{0.5cm} \\
1 & \centering \tx{if} \quad q=n-p \text { and } k=-(n+1) &  \vspace{0.5cm} \\
\left(\begin{array}{c}
-k-n-2 \\
p
\end{array}\right)\left(\begin{array}{c}
-k-p-1 \\
n-p
\end{array}\right) & \tx{if} \quad q=n\text { and } k \leqslant-(n+p+2) &  \vspace{0.5cm} \\
0 & \text { otherwise } & \quad .
\end{array}\right. 
\end{align}
It is evident that all these formulae can be easily implemented on a computer. Calculating the ranks of the quantities above is the first step of the algorithm. From now on, by a slight abuse of notation, we will indicate the groups with their dimensions, since it makes it easier to present our computations. With this in mind, the spectral sequence associated with the normal bundle of the five-fold described by \eqref{eq:Ex_configur_mat} reads:
\begin{table}[H]
\centering
\begin{tabular}{cccc|c}
    \multicolumn{4}{c|}{$E^{j,k}_1(\mathcal{E})$} & $ h^{\bullet}(\mathcal{M}, \mathcal{E})$\\
    \hline
    515\tikzmark{a1}4\tikzmark{E11}  &  \,  \tikzmark{E12}3437 & \,\, 0  \,\, & \,\, 0 \,\, &  \\
    0 \tikzmark{a2} &   0 & 0 & 0 &  \\
    0\tikzmark{a3} & 0 & 0 & 0 &   \\
    0\tikzmark{a4} & 0 & 0 & \tikzmark{b1}0 & \tikzmark{c1}1717  \\
   0\tikzmark{a5} & 0 & 0 & \tikzmark{b2}0 & \tikzmark{c2}0 \\
    0\tikzmark{a6}  & 0 & 0 & \tikzmark{b3}0 & \tikzmark{c3}0 \\
    0 & 2 & 0 & \tikzmark{b4}0 & \tikzmark{c4}0 \\
    0 & 0 & 0 & \tikzmark{b5}0 & \tikzmark{c5}0 \\
     0 & 0 & 0 & \tikzmark{b6}0 & \,\,\,\,\,\, \tikzmark{c6}2 \,\,\,. 
  \end{tabular}
    \begin{tikzpicture}[overlay, remember picture, shorten >=.5pt, shorten <=.5pt, transform canvas={yshift=.25\baselineskip}]
    \draw [->,dashed, gray] ([yshift=.75pt]{pic cs:a1}) -- ({pic cs:b1});
    \draw [->,dashed, gray] ([yshift=.75pt]{pic cs:b1}) -- ({pic cs:c1});
    \draw [->,dashed, gray] ([yshift=.75pt]{pic cs:a2}) -- ({pic cs:b2});
    \draw [->,dashed, gray] ([yshift=.75pt]{pic cs:b2}) -- ({pic cs:c2});
    \draw [->,dashed, gray] ([yshift=.75pt]{pic cs:a3}) -- ({pic cs:b3});
    \draw [->,dashed, gray] ([yshift=.75pt]{pic cs:b3}) -- ({pic cs:c3});
    \draw [->,dashed, gray] ([yshift=.75pt]{pic cs:a4}) -- ({pic cs:b4});
    \draw [->,dashed, gray] ([yshift=.75pt]{pic cs:b4}) -- ({pic cs:c4});
    \draw [->,dashed, gray] ([yshift=.75pt]{pic cs:a5}) -- ({pic cs:b5});
    \draw [->,dashed, gray] ([yshift=.75pt]{pic cs:b5}) -- ({pic cs:c5});
    \draw [->,dashed, gray] ([yshift=.75pt]{pic cs:a6}) -- ({pic cs:b6});
    \draw [->,dashed, gray] ([yshift=.75pt]{pic cs:b6}) -- ({pic cs:c6});
    \draw [->, thick] ([yshift=.75pt]{pic cs:E12}) -- ([yshift=.75pt]{pic cs:E11});
    \end{tikzpicture} 
\end{table}
It is evident that it degenerates at the first page. In fact, this is not always the case, as shown by the spectral sequence for the tangent bundle:
\begin{table}[H]
\centering
\begin{tabular}{cccc|cccc|c}
    \multicolumn{4}{c|}{$E^{j,k}_1(\mathcal{T}_{\mathcal{A}|_\mathcal{M}})$} & \multicolumn{4}{c|}{$E^{j,k}_2(\mathcal{T}_{\mathcal{A}|_\mathcal{M}})$} & $h^{\bullet} \left( \mathcal{M},\mathcal{T}_{\mathcal{A}|_\mathcal{M}}\right) $ \\
    \hline
    5\tikzmark{xA1}6\tikzmark{xE1} \, &  \tikzmark{xE2}6 & \,\, 0  \,\, & 0  & 5\tikzmark{A1}0  &  \, 0 & \,\, 0  \,\, &  0  & \\
    0\tikzmark{xA2} & 0 & 0 & 0 & 0\tikzmark{A2} & 0 & 0 & 0 &    \\
    0\tikzmark{xA3} & 0 & 0 & 0 &  0\tikzmark{A3} & 0 & 0 & 0 &  \\
   0\tikzmark{xA4} & 0 & 0 & \tikzmark{xB1}0 & 0\tikzmark{A4} & 0 & 0 & \tikzmark{B1}0 & \tikzmark{C1}50 \\
    0\tikzmark{xA5}  & 0 & 0 & \tikzmark{xB2}0 &  0\tikzmark{A5}  & 0 & 0 & \tikzmark{B2}0 & \tikzmark{C2}0 \\
    0\tikzmark{xA6} & 0 & 0 & \tikzmark{xB3}0 & 0\tikzmark{A6} & 0 & 0 & \tikzmark{B3}0 & \tikzmark{C3}0  \\
    0 & 3\tikzmark{xE} & 0 & \tikzmark{xB4}0 &  0 & 3\tikzmark{E} & 0 & \tikzmark{B4}0 & \tikzmark{C4}0\\
     0 & 0 & 0 & \tikzmark{xB5}2 & 0 & 0 & 0 & \tikzmark{B5}2 & \tikzmark{C5}$x$   \\
      0 & 0 & 0 & \tikzmark{xB6}0 & 0 & 0 & 0 & \tikzmark{B6}0 & \,\,\,\,\,\, \tikzmark{C6}$x+1$ \,\,\, . \\
  \end{tabular}
  \begin{tikzpicture}[overlay, remember picture, shorten >=.5pt, shorten <=.5pt, transform canvas={yshift=.25\baselineskip}]
    %\draw [->] ({pic cs:a}) [bend left] to ({pic cs:b});
    \draw [->,dashed, gray] ([yshift=.75pt]{pic cs:A1}) -- ({pic cs:B1});
    \draw [->,dashed, gray] ([yshift=.75pt]{pic cs:B1}) -- ({pic cs:C1});
    \draw [->,dashed, gray] ([yshift=.75pt]{pic cs:A2}) -- ({pic cs:B2});
    \draw [->,dashed, gray] ([yshift=.75pt]{pic cs:B2}) -- ({pic cs:C2});
    \draw [->,dashed, gray] ([yshift=.75pt]{pic cs:A3}) -- ({pic cs:B3});
    \draw [->,dashed, gray] ([yshift=.75pt]{pic cs:B3}) -- ({pic cs:C3});
    \draw [->,dashed, gray] ([yshift=.75pt]{pic cs:A4}) -- ({pic cs:B4});
    \draw [->,dashed, gray] ([yshift=.75pt]{pic cs:B4}) -- ({pic cs:C4});
    \draw [->,dashed, gray] ([yshift=.75pt]{pic cs:A5}) -- ({pic cs:B5});
    \draw [->,dashed, gray] ([yshift=.75pt]{pic cs:B5}) -- ({pic cs:C5});
    \draw [->,dashed, gray] ([yshift=.75pt]{pic cs:A6}) -- ({pic cs:B6});
    \draw [->,dashed, gray] ([yshift=.75pt]{pic cs:B6}) -- ({pic cs:C6});
    \draw [->, thick] ([yshift=.75pt]{pic cs:B5}) -- ({pic cs:E});
    \draw [->, thick] ([yshift=.75pt]{pic cs:xE2}) -- ([yshift=.75pt]{pic cs:xE1});
  \end{tikzpicture} 
\end{table}
We see that this sequence has a non-trivial map in it which prevents it from degenerating. By defining $x$ as the kernel of such a map (so that, in this case, $x \in \{0,1,2 \})$, we can proceed to the subsequent pages, and obtain the cohomology on the right. We can plug these results into the long cohomology sequence associated to the adjunction sequence, i.e. \eqref{eq:Long_adjunction_sequence}. Then, by indicating the cohomologies with their dimensions, we have the following schematic sequence:
\begin{align}
\begin{aligned}
 0 \xrightarrow[]{} 0  &\rightarrow 50  \stackrel{}{\rightarrow} 1717 \rightarrow \\
 h^{4,1}  &  \stackrel{}{\rightarrow} 0 \stackrel{}{\rightarrow} 0 \rightarrow \\
h^{3,1} &  \stackrel{}{\rightarrow} 0 \stackrel{}{\rightarrow} 0 \rightarrow \\
h^{2,1} 
&  \stackrel{}{\rightarrow} 0 \stackrel{}{\rightarrow} 0 \rightarrow \\
h^{1,1} 
&  \stackrel{}{\rightarrow} x \stackrel{}{\rightarrow} 0 \rightarrow \\
0 &  \stackrel{}{\rightarrow} x+1 \stackrel{}{\rightarrow} 2 \xrightarrow[]{} 0 . 
\end{aligned}
\end{align}
Clearly, the associated system of equations is:
\begin{align}
    x+1 -2 =0 \, , \quad h^{1,1} = x \, , \quad 50 - 1717 + h^{4,1} = 0,
\end{align}
with the other Hodge numbers being zero. 

Let us now move to the computations for the symmetrised adjunction sequence. Once again, we have some spectral sequences to consider, but let us employ a couple of shortcuts to write them more compactly. We avoid drawing the horizontal maps in the first row just for aesthetic reasons, while we explicitly draw arrows to represent all the other maps. We also do not write out all the elements of the subsequent pages, like we did for the tangent bundle, since they follow straightforwardly from the first page. With this in mind, the spectral sequences involved in the calculation of the other Hodge numbers schematically read: \\
\begin{table}[H]
\centering
\begin{tabular}{cccc|c}
    \multicolumn{4}{c|}{$E^{j,k}_1( \mathrm{Sym}^2 \mathcal{E})$} & $h^{\bullet}(\mathcal{M}, \mathrm{Sym}^2 \mathcal{E})$\\
    \hline
    253\tikzmark{aa1}170  &  \, 258324 & \, 45631   & \,\, 0 \,\, &  \\
    0 \tikzmark{aa2} &   0 & 0 & 0 &  \\
    0\tikzmark{aa3} & 0 & 0 & 0 &   \\
    0\tikzmark{aa4} & 0 & 0 & \tikzmark{ab1}0 & \tikzmark{ac1}40477 \\
    0\tikzmark{aa5}  & 0 & 0 & \tikzmark{ab2}0 & \tikzmark{ac2}0 \\
    0\tikzmark{aa6} & 0 & 0 & \tikzmark{ab3}0 & \tikzmark{ac3}0 \\
    0 & 9\tikzmark{aE2} & \tikzmark{aE1}6 & \tikzmark{ab4}0 & \tikzmark{ac4}0 \\
     0 & 0 & 0 & \tikzmark{ab5}0 & \tikzmark{ac5}$x$ \\
      0 & 0 & 0 & \tikzmark{ab6}0 & \,\,\,\,\,\, \tikzmark{ac6}$x+3$ \,\,\, , \\
  \end{tabular}
\begin{tikzpicture}[overlay, remember picture, shorten >=.5pt, shorten <=.5pt, transform canvas={yshift=.25\baselineskip}]
    %\draw [->] ({pic cs:a}) [bend left] to ({pic cs:b});
    \draw [->,dashed, gray] ([yshift=.75pt]{pic cs:aa1}) -- ({pic cs:ab1});
    \draw [->,dashed, gray] ([yshift=.75pt]{pic cs:ab1}) -- ({pic cs:ac1});
    \draw [->,dashed, gray] ([yshift=.75pt]{pic cs:aa2}) -- ({pic cs:ab2});
    \draw [->,dashed, gray] ([yshift=.75pt]{pic cs:ab2}) -- ({pic cs:ac2});
    \draw [->,dashed, gray] ([yshift=.75pt]{pic cs:aa3}) -- ({pic cs:ab3});
    \draw [->,dashed, gray] ([yshift=.75pt]{pic cs:ab3}) -- ({pic cs:ac3});
    \draw [->,dashed, gray] ([yshift=.75pt]{pic cs:aa4}) -- ({pic cs:ab4});
    \draw [->,dashed, gray] ([yshift=.75pt]{pic cs:ab4}) -- ({pic cs:ac4});
    \draw [->,dashed, gray] ([yshift=.75pt]{pic cs:aa5}) -- ({pic cs:ab5});
    \draw [->,dashed, gray] ([yshift=.75pt]{pic cs:ab5}) -- ({pic cs:ac5});
    \draw [->,dashed, gray] ([yshift=.75pt]{pic cs:aa6}) -- ({pic cs:ab6});
    \draw [->,dashed, gray] ([yshift=.75pt]{pic cs:ab6}) -- ({pic cs:ac6});
    \draw [->, thick] ([yshift=.75pt]{pic cs:aE1}) -- ([yshift=.75pt]{pic cs:aE2});
\end{tikzpicture} 
\end{table}
\begin{table}[H]
\centering
\begin{tabular}{cccc|c}
    \multicolumn{4}{c|}{$E^{j,k}_1(\mathcal{T}_{\mathcal{A}}|_{\mathcal{M}} \otimes \mathcal{E})$} & $h^{\bullet}\left(\mathcal{M},  \mathcal{T}_{\mathcal{A}}|_{\mathcal{M}} \otimes \mathcal{E}\right)$ \\
    \hline
    8397\tikzmark{0a1}3  &  \, 66346 & 5160 & \,\, 0 \,\, &  \\
    0 \tikzmark{0a2} &   0 & 0 & 0 &  \\
    0\tikzmark{0a3} & 0 & 0 & 0 &   \\
    0\tikzmark{0a4} & 0 & 0 & \tikzmark{0b1}0 & \tikzmark{0c1}22787  \\
    0\tikzmark{0a5}  & 0 & 0 & \tikzmark{0b2}0 & \tikzmark{0c2}0 \\
    0\tikzmark{0a6} & 2 & 0 & \tikzmark{0b3}0 & \tikzmark{0c3}0 \\
    0 & 16\tikzmark{0E2} & \tikzmark{0E1}12 & \tikzmark{0b4}0 & \tikzmark{0c4}0 \\
     0 & 0 & 0 & \tikzmark{0b5}0 & \tikzmark{0c5}$y+2$ \\
      0 & 0 & 0 & \tikzmark{0b6}0 & \,\,\,\,\,\, \tikzmark{0c6}$y+4$ \,\,\, , \\
\end{tabular}
\begin{tikzpicture}[overlay, remember picture, shorten >=.5pt, shorten <=.5pt, transform canvas={yshift=.25\baselineskip}]
    %\draw [->] ({pic cs:a}) [bend left] to ({pic cs:b});
    \draw [->,dashed, gray] ([yshift=.75pt]{pic cs:0a1}) -- ({pic cs:0b1});
    \draw [->,dashed, gray] ([yshift=.75pt]{pic cs:0b1}) -- ({pic cs:0c1});
    \draw [->,dashed, gray] ([yshift=.75pt]{pic cs:0a2}) -- ({pic cs:0b2});
    \draw [->,dashed, gray] ([yshift=.75pt]{pic cs:0b2}) -- ({pic cs:0c2});
    \draw [->,dashed, gray] ([yshift=.75pt]{pic cs:0a3}) -- ({pic cs:0b3});
    \draw [->,dashed, gray] ([yshift=.75pt]{pic cs:0b3}) -- ({pic cs:0c3});
    \draw [->,dashed, gray] ([yshift=.75pt]{pic cs:0a4}) -- ({pic cs:0b4});
    \draw [->,dashed, gray] ([yshift=.75pt]{pic cs:0b4}) -- ({pic cs:0c4});
    \draw [->,dashed, gray] ([yshift=.75pt]{pic cs:0a5}) -- ({pic cs:0b5});
    \draw [->,dashed, gray] ([yshift=.75pt]{pic cs:0b5}) -- ({pic cs:0c5});
    \draw [->,dashed, gray] ([yshift=.75pt]{pic cs:0a6}) -- ({pic cs:0b6});
    \draw [->,dashed, gray] ([yshift=.75pt]{pic cs:0b6}) -- ({pic cs:0c6});
    \draw [->,thick] ([yshift=.75pt]{pic cs:0E1}) -- ([yshift=.75pt]{pic cs:0E2});
\end{tikzpicture}
\end{table}
and
\begin{table}[H]
\centering
\begin{tabular}{cccc|c}
    \multicolumn{4}{c|}{$E^{j,k}_1(\mathcal{T}_{\mathcal{A}}|_{\mathcal{M}}^{\,\,2})$} & $H^{\bullet}\left(\mathcal{M}, \mathcal{T}_{\mathcal{A}}|_{\mathcal{M}}^{\,\,2}\right)$ \\
    \hline
    93\tikzmark{1A1}4  &  \, 300 & 3  & \,\, 0 \,\, &  \\
    0 \tikzmark{1A2} &   0 & 0 & 0 &  \\
    0\tikzmark{1A3} & 0 & 0 & 0 &   \\
    0\tikzmark{1A4} & 0 & 0 & \tikzmark{1B1}0 & \tikzmark{1C1}637  \\
    0\tikzmark{1A5}  & 0 & 0 & \tikzmark{1B2}0 & \tikzmark{1C2}0 \\
    0\tikzmark{1A6} & 3\tikzmark{14E} & 0 & \tikzmark{1B3}0 & \tikzmark{1C3}0 \\
    0 & 6\tikzmark{11E} & \tikzmark{12E}6 & \tikzmark{1B4}3 & \tikzmark{1C4}$b$ \\
     0 & 0 & 0 & \tikzmark{1B5}0 & \tikzmark{1C5}$b +3 -z +a$ \\
      0 & 0 & 0 & \tikzmark{1B6}0 & \,\,\,\,\,\, \tikzmark{1C6}$a +3 - z$ \,\,\, . \\
\end{tabular}
\begin{tikzpicture}[overlay, remember picture, shorten >=.5pt, shorten <=.5pt, transform canvas={yshift=.25\baselineskip}]
    %\draw [->] ({pic cs:a}) [bend left] to ({pic cs:b});
    \draw [->,dashed, gray] ([yshift=.75pt]{pic cs:1A1}) -- ({pic cs:1B1});
    \draw [->,dashed, gray] ([yshift=.75pt]{pic cs:1B1}) -- ({pic cs:1C1});
    \draw [->,dashed, gray] ([yshift=.75pt]{pic cs:1A2}) -- ({pic cs:1B2});
    \draw [->,dashed, gray] ([yshift=.75pt]{pic cs:1B2}) -- ({pic cs:1C2});
    \draw [->,dashed, gray] ([yshift=.75pt]{pic cs:1A3}) -- ({pic cs:1B3});
    \draw [->,dashed, gray] ([yshift=.75pt]{pic cs:1B3}) -- ({pic cs:1C3});
    \draw [->,dashed, gray] ([yshift=.75pt]{pic cs:1A4}) -- ({pic cs:1B4});
    \draw [->,dashed, gray] ([yshift=.75pt]{pic cs:1B4}) -- ({pic cs:1C4});
    \draw [->,dashed, gray] ([yshift=.75pt]{pic cs:1A5}) -- ({pic cs:1B5});
    \draw [->,dashed, gray] ([yshift=.75pt]{pic cs:1B5}) -- ({pic cs:1C5});
    \draw [->,dashed, gray] ([yshift=.75pt]{pic cs:1A6}) -- ({pic cs:1B6});
    \draw [->,dashed, gray] ([yshift=.75pt]{pic cs:1B6}) -- ({pic cs:1C6});
    \draw [->, thick] ([yshift=-0.75pt]{pic cs:1B4}) -- ([yshift=-0.75pt]{pic cs:11E});
    \draw [->, thick] ([yshift=1.75pt]{pic cs:12E}) -- ([yshift=1.75pt]{pic cs:11E});
    \draw [->, thick] ([yshift=-0.75pt]{pic cs:1B4}) -- ([yshift=-0.75pt]{pic cs:14E});
\end{tikzpicture} 
\end{table}
In all three cases there are maps preventing the sequence from degenerating at the first page, which are taken into account as shown. This introduces additional unknowns in the problem. Plugging our results in sequences \eqref{eq:Symmetrised_cohom_sequence_1} and \eqref{eq:Symmetrised_cohom_sequence_2}, and focusing on the dimensions as above, we obtain:
\begin{align}
\begin{aligned}
0 \xrightarrow[]{} 0 &\rightarrow   637 \stackrel{}{\rightarrow} k_6 \rightarrow  \\
0 &  \stackrel{}{\rightarrow} 0 \stackrel{}{\rightarrow}k_5 \rightarrow \\
h^{2,3}
&  \stackrel{}{\rightarrow} 0 \stackrel{}{\rightarrow} k_4 \rightarrow \\
h^{2,2}
&  \stackrel{}{\rightarrow} b \stackrel{}{\rightarrow} k_3 \rightarrow \\
0
&  \stackrel{}{\rightarrow}b +3 - z + a \stackrel{}{\rightarrow} k_2 \rightarrow
\\
0 
&  \stackrel{}{\rightarrow} a + 3 - z \stackrel{}{\rightarrow} k_1 \xrightarrow[]{} 0
,
\label{eq:Symmetrised_cohom_sequence_ex_1}
\end{aligned}
\end{align}
and
\begin{align}
\begin{aligned}
0 \xrightarrow[]{} k_6 &\rightarrow   22787 \stackrel{}{\rightarrow} 40477 \rightarrow \\
k_5 &  \stackrel{}{\rightarrow} 0 \stackrel{}{\rightarrow}0 \rightarrow \\
k_4
&  \stackrel{}{\rightarrow} 0 \stackrel{}{\rightarrow} 0 \rightarrow \\
k_3
&  \stackrel{}{\rightarrow} 0 \stackrel{}{\rightarrow} 0 \rightarrow \\
k_2 
&  \stackrel{}{\rightarrow} y+2 \stackrel{}{\rightarrow} x \rightarrow
\\
k_1 
&  \stackrel{}{\rightarrow} y+4 \stackrel{}{\rightarrow} x+3 \xrightarrow[]{} 0.
\label{eq:Symmetrised_cohom_sequence_ex_2}
\end{aligned}
\end{align}
We find 5 non-trivial equations from \eqref{eq:Symmetrised_cohom_sequence_ex_1} and 4 non-trivial equations from \eqref{eq:Symmetrised_cohom_sequence_ex_2}. The system can be enlarged by adding the Euler equation (see \eqref{Euler0}) and the constraint coming from the index theorem discussed (see \eqref{Hodgecons}).\footnote{The Euler number is found, with a straightforward implementation of the calculation described at the end of section \ref{sec:Basic_properties}, to be $-39984$. } Moreover, it easy to see that the auxiliary variables $x$ and $y$ drop out of the system, leaving a system for which a unique solution can be found.
Ignoring the supplementary variables, the result of this computation yields the Hodge diamond:
\begin{equation}
 \begin{array}{ccccccccccc}
           &          &          &          &          &     1   &          &          &          &          &            \\
           &          &          &          &    0    &          &   0     &         &          &           &            \\
           &          &          &    0    &          &1&        &   0    &          &          &             \\ 
           &          &   0    &           &0&          &0&     &  0      &         &              \\     
           &  0      &         &0&          &1&        &0&     &     0    &              \\ 
    1    &          &1667&          &18327&          &18327&        &1667&    &    1     \\
          &  0      &         &0&          &1&        &0&     &     0    &              \\ 
           &          &   0    &           &0&          &0&     &  0      &         &              \\     
           &          &          &    0    &          &1&        &   0    &          &          &             \\ 
           &          &          &          &    0    &          &   0     &         &          &           &            \\
           &          &          &          &          &     1   &          &          &          &          & .
 \end{array} 
 \label{hodgediamond}
\end{equation}

\addcontentsline{toc}{section}{References}
\bibliography{mergedbiblio.bib}

\end{document}